%% file: microcoupler_main.tex
\documentclass{iopjournal}
\usepackage{lmodern}
\usepackage{longtable}
\usepackage{amsmath}
\usepackage{amssymb}
\usepackage{multirow}
\usepackage[table]{xcolor}
\usepackage{graphicx}
\usepackage{url}

\renewcommand{\headrulewidth}{0pt}

\definecolor{pmgrey}{gray}{0.90}
\definecolor{nearpmgrey}{gray}{0.78}

\begin{document}


\title{Polarisation-resolved identification of spontaneous four-wave mixing processes in a multimode fused tapered fibre coupler}

\author{Jefferson Fl\'orez${^1}{^\dagger}$, Chams Baker${^1}{^\ddagger}$, Benjamin J. Sussman$^2$, Xiaoyi Bao$^1$, Jeff S. Lundeen$^3$ and Lambert Giner$^{4,*}$}

\affil{$^1$ Department of Physics, University of Ottawa, Ottawa, Canada}
\affil{$^2$ National Research Council of Canada, Ottawa, Canada}
\affil{$^3$ Nexus for Quantum Technologies, Department of Physics, University of Ottawa, Ottawa, Canada}
\affil{$^4$ D\'epartement de physique et astronomie, Universit\'e de Moncton, Moncton, Canada}
\affil{$^\dagger$ Current affiliation: Quantinuum, Cambridge, United Kingdom}
\affil{$^\ddagger$ Current affiliation: Corning Incorporated, Corning, USA}
\affil{$^{*}$ Author to whom any correspondence should be addressed.}
\email{lambert.giner@umoncton.ca}

\begin{abstract}
Integrated quantum photonics benefits from photon-pair sources that generate photons directly in waveguides, where they can be efficiently collected, routed, and manipulated. Here, we investigate spontaneous four-wave mixing (SFWM) in a fused tapered-fibre microcoupler formed from two single-mode fibres, and report four contributions. First, we observe SFWM photon pairs in this multimode device: its elliptical central region supports three spatial mode profiles, each with two polarisations, giving six guided modes and allowing many distinct phase-matching processes, and pumping it near 800~nm yields two photon pairs at 648/1048~nm and 660/1021~nm. Second, because several theoretically allowed SFWM processes produce similar wavelengths and wavelength alone therefore does not identify their origin, we distinguish the processes by combining phase-matching calculations and selection rules with two independent polarisation measurements: the dependence of the coincidence rate on pump polarisation and polarisation tomography of the generated photons. Third, within the present geometrical model, we find that the 660/1021 nm pair is generated by co-polarised pump photons in a process occurring entirely within a single higher-order mode, whereas the 648/1048~nm pair is generated by orthogonally polarised pump photons in an intermodal process coupling a fundamental and a higher-order mode; the signal photons from the two processes have similar output polarisations, while the idler photons are nearly orthogonal. Fourth, extending the analysis beyond the measured operating point, we theoretically identify pairs of simultaneous SFWM processes driven by a common pump that could generate either polarisation entanglement or composite spatial-polarisation entanglement, depending on the pump wavelength. More generally, this work provides a practical strategy for identifying intermodal SFWM processes in multimode waveguides when spectral information alone is insufficient.
\end{abstract}


\input{section1_introduction}

\input{section2_device}

\input{section3_experiment}

\input{section4_modelling}

\input{section5_outlook}


\input{section6_conclusion}

\bibliographystyle{iopart-num}
\bibliography{references}

\section*{Acknowledgments}
This work was supported in part by the Canada Foundation for Innovation (CFI), the Natural
Sciences and Engineering Research Council of Canada (NSERC) and the Canada Research Chairs
program. We warmly thank Duncan England and Philip J. Bustard (National Research Council of
Canada) for their experimental support and for many helpful discussions and advice.

During the preparation of this manuscript, the authors used Claude
(Anthropic, Opus models 4.8 to 5.5) to edit human-written text, including
improvements to language, clarity and structure, to generate figures from
data produced by the authors, and to support the literature review. All
references were checked by the authors against the original sources. The
authors critically revised all output and take full responsibility for the
accuracy, integrity and originality of the content.

\section*{Data availability statement}
The data that support the findings of this study are available upon reasonable request from the authors.


\end{document}

%% file: section1_introduction.tex
\section{Introduction}
\label{sec:introduction}

Photonic quantum technologies, from quantum communication to quantum information, rely on
sources of photon pairs~\cite{GisinThew2007,Pirandola2020,OBrien2009,Flamini2019}, and a
subset of these applications further requires the two photons of a pair to be
entangled~\cite{Horodecki2009}. Such pairs are commonly generated by spontaneous parametric
down-conversion (SPDC), in which a pump photon spontaneously generates a pair of photons.
Bulk nonlinear-crystal SPDC has yielded efficient free-space
sources~\cite{Anwar2021}. For integrated applications such as fibre quantum networks or
photonic circuits, however, the photons emitted by these bulk crystals must subsequently be
coupled into a single-mode fibre or waveguide. While this coupling can be efficient in
practice~\cite{Guerreiro2013}, it demands precise mode engineering and suffers from a
trade-off between brightness (the number of pairs generated per second) and heralding
efficiency (the probability of finding a photon in the fibre once its twin has been
detected)~\cite{Bennink2010}. Increasing one generally comes at the expense of the other.
In contrast, a waveguide source of photon pairs is intrinsically
compatible with integrated optics. This manuscript experimentally and theoretically
investigates a tapered optical-fibre waveguide with an elliptical cross-section as a source
of photon pairs.

While SPDC can be implemented in waveguides, it is restricted to the relatively small set of
waveguide platforms that can be fabricated from materials with a second-order nonlinearity
($\chi^{(2)}$)~\cite{Martin2010}. Its third-order nonlinear counterpart, spontaneous
four-wave mixing (SFWM), occurs in a wider range of materials~\cite{Afsharnia2024} and is therefore compatible
with more types of waveguides, including standard silica optical fibres. In SFWM, two pump
photons spontaneously convert into a pair of photons. Fibres have two advantages for SFWM:
they confine the light to a small area, increasing the intensity, and they can provide long
interaction lengths, both of which increase the probability of generating a photon pair. On
the other hand, simultaneously satisfying energy and momentum conservation between the pump
photons and the generated pair---that is, achieving phase matching---can be more challenging.
In optical fibres, phase matching is typically achieved using birefringent or intermodal
processes~\cite{GarayPalmett2023,garaypalmett2016,rottwitt2018,majchrowska2022,gawlik2025,shahar2023}. The fibre
photon-pair source studied here further enhances the nonlinear interaction by tapering the
fibre, thereby reducing the effective mode area~\cite{Shukhin2020}.

An ideal integrated photon-pair source would emit the two photons directly into separate
waveguides, so that they are deterministically separated and can be independently routed and
detected without lossy post-generation separation. In addition, because the waveguide itself
would provide a distinct label for each photon, the photons' other degrees of freedom, such as
polarisation and frequency, remain available for encoding entanglement. Our source naturally
incorporates two input and two output waveguides. Known as a microcoupler, it is fabricated by
flame-tapering two single-mode fibres held in contact and fused over a central region. In this
region, the microcoupler has an elliptical cross-section with transverse dimensions on the
order of a micrometre. It supports three spatial mode profiles---LP$_{01}$,
LP$_{11}$-even, and LP$_{11}$-odd---each with two polarisations, $x$ and $y$, giving six
guided modes in total. We pump the microcoupler near 800~nm with a femtosecond pulsed laser.
The multiple spatial modes and polarisations allow many distinct SFWM processes and therefore
a rich set of possible phase-matching conditions.

In a multimode waveguide, identifying the SFWM process responsible for a detected photon pair
can be difficult. Each field can occupy different spatial modes and polarisations, giving many
possible phase-matched processes, some of which produce photons at similar wavelengths.
Previous work has distinguished such processes either by designing them to be spectrally
separated~\cite{CruzDelgado2016} or by comparing measured wavelengths with calculated
phase-matching conditions~\cite{Montazeri2024}. When several processes remain compatible with
the measured wavelengths, additional information is required. Here, we begin with all
spatial-mode and polarisation combinations theoretically allowed in the microcoupler, use phase
matching to eliminate most of them, and then use two independent polarisation
measurements---the pump-polarisation dependence of the coincidence rate and polarisation
tomography of the generated photons---to narrow the remaining candidates. A final comparison
of the phase-matching behaviour resolves the residual ambiguity: the combined
observations favour a single pair of processes for the observed photon pairs within the
present geometrical model. In this way, we describe a general
identification strategy, combining selection rules, phase-matching calculations, and
independent experimental signatures, that researchers can use to determine which nonlinear
processes occur in a multimode waveguide when spectral information alone is insufficient.

Finally, we extend the analysis beyond the identified processes. Leaving the pump wavelength
as a free parameter, we predict theoretically that the same microcoupler would allow two simultaneous processes
driven by a common pump wavelength to generate entangled photons. Depending on the chosen
operating point, this entanglement may reside in polarisation alone or, in a composite manner,
in both spatial mode and polarisation. The realisation of such a source remains conditioned by
the geometry of the microcoupler, whose ellipticity determines how well the spatial modes are
separated.

The contributions of this work are thus fourfold: the observation of SFWM photon pairs in a
multimode fused microcoupler; an identification method combining selection rules,
phase-matching calculations and two independent polarisation measurements; the resulting
attribution, within the present geometrical model, of the two observed pairs to an intramodal
and an intermodal process; and the theoretical prediction of simultaneous processes that could
generate polarisation or composite entanglement. This article is organised as follows.
Section~\ref{sec:device} describes the microcoupler, its fabrication and the modes guided in
its central region. Section~\ref{sec:experiment} presents the experimental setup and the two
polarisation measurements. Section~\ref{sec:modelling} details the modelling and
identification of the processes. Section~\ref{sec:outlook} finally extends the analysis to the
generation of polarisation-entangled states.

%% file: section2_device.tex
\section{Device}
\label{sec:device}

The photon pairs studied in this work are generated in a microcoupler, a fused tapered-fibre device whose narrow, elliptical and multimode central region is key to what follows. We describe its fabrication, then its structure and guided modes.

\subsection{Fabrication}
\label{sec:fabrication}

The device is made from two single-mode fibres (Fibercore SM600) tapered
together under a flame while kept in contact, forming the composite waveguide we refer
to as a microcoupler throughout this article. These fibres operate optimally between
633 and 780~nm and have a 125~$\mu$m cladding, a numerical aperture of 0.10 to 0.14,
and a mode-field diameter of 3.6 to 5.3~$\mu$m at 633~nm; their cut-off wavelength lies
between 500 and 600~nm, so they remain single-mode at all the wavelengths used in the
experiment. During tapering, the two fibres fuse over a central region about 10~cm long, whose thinnest
section, referred to as the waist as is customary for tapered fibres~\cite{birks1992},
reaches a target diameter on the order of one micron.

Fabrication proceeds in three stages: sample preparation, tapering, and packaging to
protect the device from dust and mechanical shock. The two fibres are first cleaved,
cleaned, and mounted on the tapering rig, each end fixed to a motorised,
computer-controlled stage, and loosely wound once around each other so they stay in
contact during the pull. A hydrogen flame, also on a motorised stage, then sweeps back
and forth along the contact region, the heated span widening as the fibre ends are slowly
drawn apart, following the heat-brush approach of~\cite{baker2011,birks1992}. Throughout
the pull, a laser at 810~nm (reaching about 90\% transmission) is launched into one fibre
and its transmitted power recorded against the stage extension: as the claddings thin and
the two cores draw closer, their evanescent fields overlap and the modes couple, giving
the transmitted intensity a periodic structure that confirms the fibre has not
broken~\cite{bilodeau1987}.

The pull is stopped using a calibration of core diameter against extension, which reaches
1.05~$\mu$m at 200~mm of extension. Because the stopping criterion is the diameter rather
than a splitting-ratio value, the splitting ratio is not set in advance; once the device
is made, all input/output port combinations are characterised and the pair with the
highest transmission is retained. After the photon-pair experiments
(Sec.~\ref{sec:experiment}), the microcoupler is cleaved at several points to examine its
cross section and determine its shape and transverse dimensions.

\subsection{Structure and guided modes}
\label{sec:structure}

Having described its fabrication, we now turn to the physical structure of the
microcoupler and the spatial modes it supports. It consists of three longitudinal
sections: two transition regions, where the initially independent fibres progressively
fuse, and a central region where they merge into one; at each end, the unstretched SM600
fibres continue the device, remaining independent and single-mode and acting as input and
output ports.

\begin{figure}[htpb]
\centering
\begin{tabular}{cc}
\includegraphics[scale=0.20]{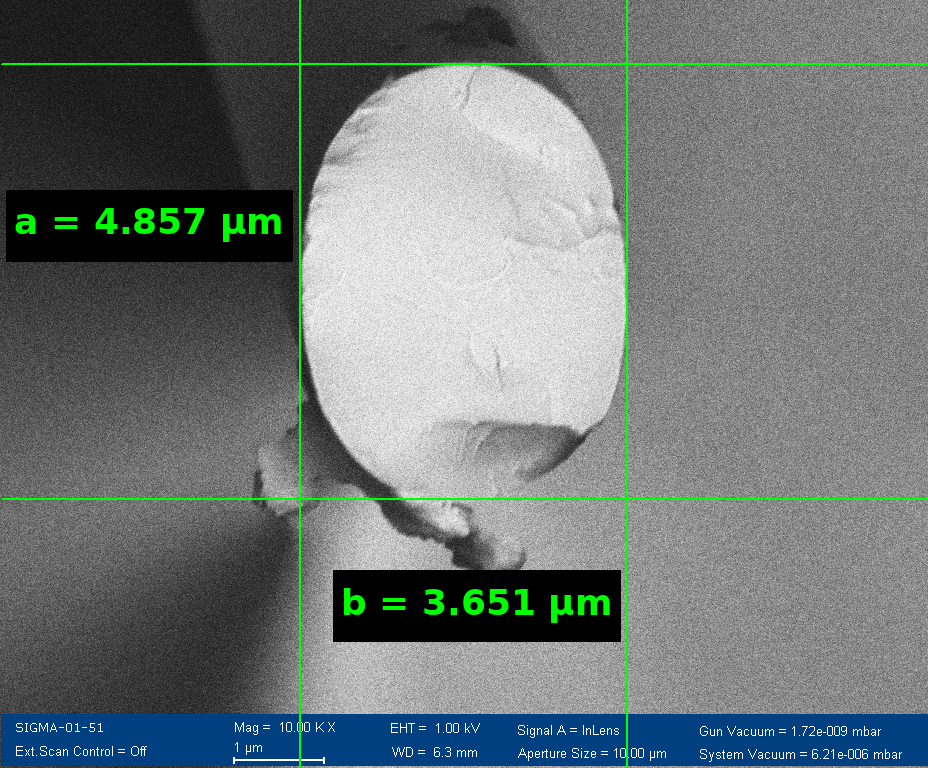} &
\includegraphics[scale=0.20]{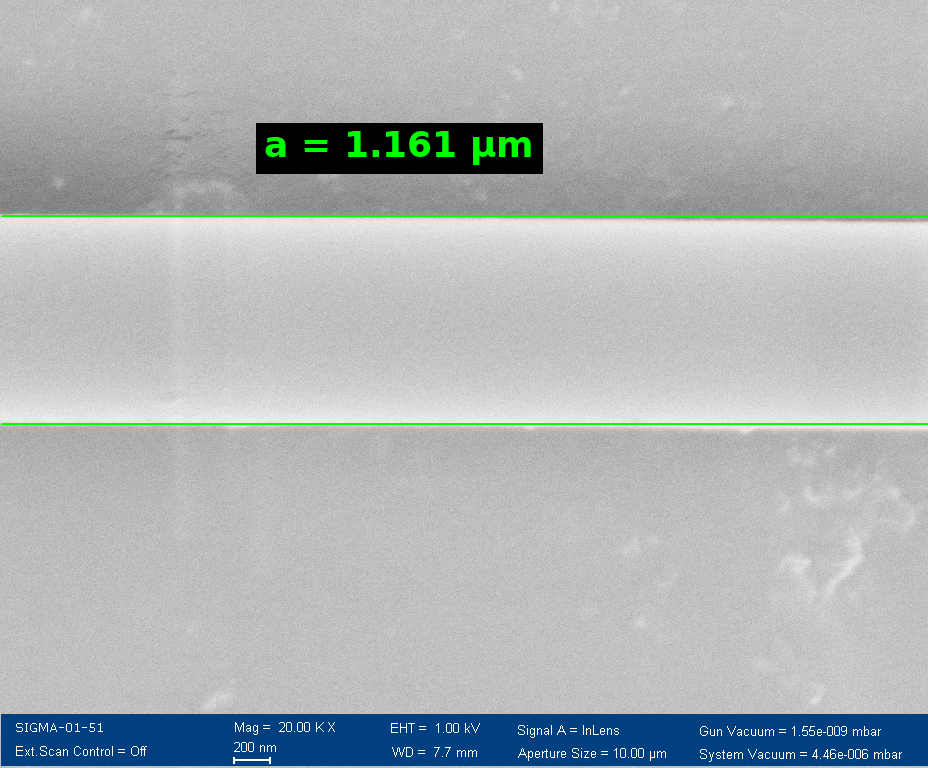}
\end{tabular}
\caption{SEM images of the microcoupler: (a)~cross section away from the
waist, used to read the axis ratio (major axis $4.857~\mu$m, minor axis
$3.651~\mu$m, ratio $\approx 1.33$); (b)~top view at the waist, used to read the
absolute scale (major axis $1.161~\mu$m). The bright region is the broken fibre
section (destructive imaging); the visible marks are fracture artefacts, not the
shape of the guide.}
\label{fig:sem_pics}
\end{figure}

The central region is modelled as an elliptical fibre~\cite{wang2005} (see figure~\ref{fig:sem_pics}) whose dimensions and
ellipticity, given in Sec.~\ref{sec:modelling}, allow six modes to propagate, with the
transverse profiles shown in figure~\ref{fig:modes_profiles}. These comprise the two
fundamental LP$_{01}$ modes (LP$_{01}^{x}$ and LP$_{01}^{y}$) and four first-excited
LP$_{11}$ modes (LP$_{11}^{x\text{-}\mathrm{even}}$, LP$_{11}^{y\text{-}\mathrm{even}}$,
LP$_{11}^{x\text{-}\mathrm{odd}}$, and LP$_{11}^{y\text{-}\mathrm{odd}}$)~\cite{keiser}, where $x$ and
$y$ denote the direction of the transverse electric field, that is, the polarisation, and
even and odd denote the spatial symmetry of the field about the major axis of the elliptical
core (even: vertical nodal plane; odd: horizontal nodal plane)~\cite{blackgagnon}.

Because the six modes supported in the central region have different effective indices,
nonlinear processes that exploit these differences become possible, in particular
SFWM~\cite{agrawal}. A pump injected through the input port crosses the first transition
region and generates photon pairs in the central region; these photons then cross the
second transition region and reach the fibre outputs leading to the detection system. As
a result, owing to the uncontrolled residual birefringence of the single-mode fibres, the
photons undergo unknown unitary polarisation transformations, which calls for the careful
analysis framework set out in Sec.~\ref{sec:experiment}.

%% file: section3_experiment.tex
\section{Experiment}
\label{sec:experiment}

We now turn from the device itself to the light it produces, establishing that
the photon pairs arise from SFWM and characterising their spectral and
polarisation properties.

\subsection{Source and detected pairs}
\label{sec:source}

\begin{figure}[htpb]
\centering
\includegraphics[width=\textwidth]{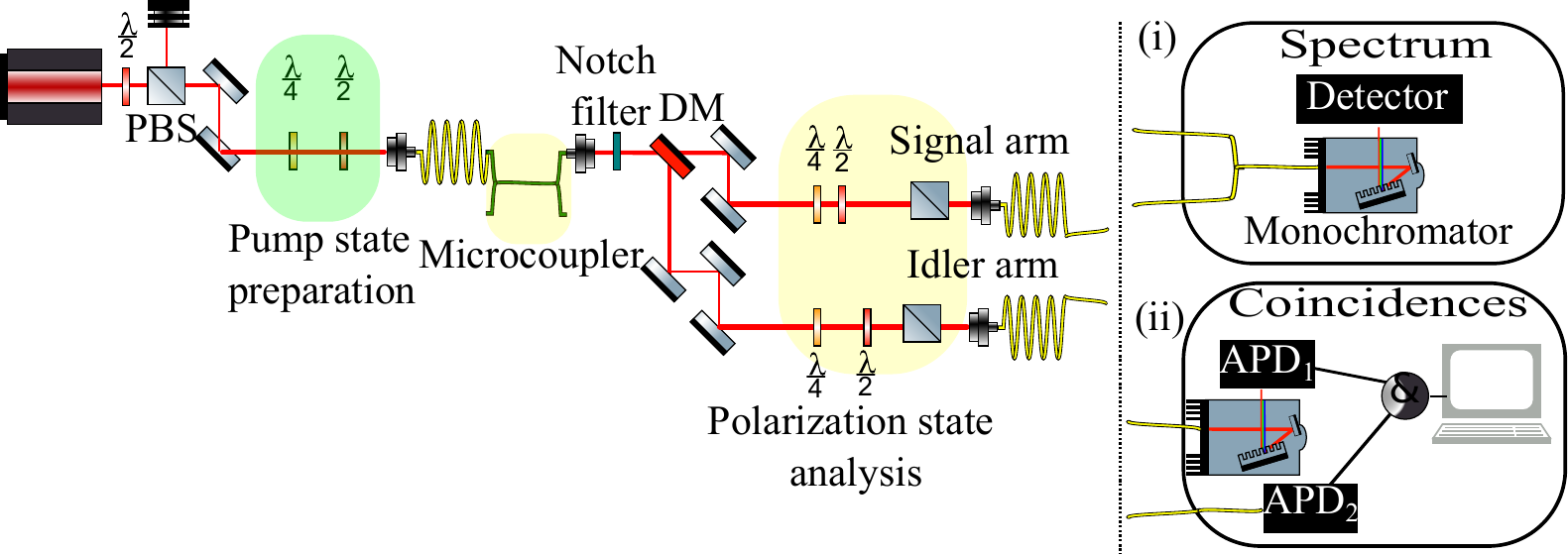}
\caption{Experimental setup. The pump is prepared by a quarter- and a half-wave
plate (QWP, HWP) before the microcoupler; immediately after the device a notch
filter rejects the pump, and a dichroic mirror then splits the output into a
signal and an idler analysis arm. Two detection configurations are used:
(i)~a spectrometer with an EMCCD camera for spectral identification, and
(ii)~a pair of avalanche photodiodes (APDs) in coincidence, one arm filtered by
the spectrometer used as a monochromator, for the $g^{(2)}(0)$ and
pump-dependence measurements; adding a polarisation analyser (QWP, HWP, PBS) on
the filtered arm realises the tomography.}
\label{fig:exp_setup}
\end{figure}

The microcoupler described in the previous section is pumped by a
titanium--sapphire laser (Spectra-Physics Tsunami) delivering Fourier-limited
$100$~fs pulses at an $80$~MHz repetition rate, centred around $800$~nm. Being
Fourier-limited, these pulses give the pump a spectrum spanning roughly $795$ to
$805$~nm, a bandwidth used throughout this article and in particular in
Section~\ref{sec:modelling}. The pump polarisation is prepared before the input
port by a half-wave plate (HWP) and a quarter-wave plate (QWP)
(see Fig.~\ref{fig:exp_setup}): the QWP compensates the ellipticity introduced
as the pump propagates through the non-polarisation-maintaining fibres, while the
HWP sweeps the resulting linear polarisation. Photon pairs are then generated by
SFWM in the central region and leave the microcoupler through one of its two
output fibres; for experimental simplicity, only the input and output ports with
the best transmission are used, as discussed in the previous section, which
introduces losses proportional to the splitting ratio but does not affect the
analysis of the results. Once in free space, a notch filter rejects the pump, and
a dichroic mirror then separates each pair into two analysis arms, one for the
signal and one for the idler. Preliminary results from this experiment were
reported in a Master's thesis~\cite{Cheng2017}.

Depending on the signals to be acquired, the setup operates in two
configurations (see Fig.~\ref{fig:exp_setup}). In configuration~(i), used for
spectral identification, both analysis arms are connected to a spectrometer. In
configuration~(ii), used for coincidence detection, one arm is filtered by the
spectrometer acting as a monochromator and sent to an APD, while the other goes
directly to an APD that heralds the conjugate photon. This configuration serves
both the $g^{(2)}(0)$ and the pump-dependence measurements; adding a
polarisation-analysis system (a QWP, a HWP, and a polarising beam splitter, PBS)
on the filtered arm turns it into the tomography setup.

In configuration~(i), the spectrometer (Princeton Instruments Acton SP2300) is
paired with an EMCCD camera (Princeton Instruments ProEM 1600) whose efficiency
is $95\%$ near $650$~nm but only about $2\%$ beyond $1050$~nm, so the signal
photons are readily identified. With about $20$~mW of pump through the
microcoupler, their spectrum is recorded over the $400$ to $800$~nm range, where
the camera efficiency is high, as the average of $50$ frames of $20$~seconds each
using a $600$~lines/mm grating, with the averaged background subtracted, and
reveals two photons, at $648$~nm and $660$~nm. These peaks are genuine: they
persist when the grating angle or the grating itself is changed, and their
intensity scales quadratically with pump power, confirming the SFWM origin.

The idlers are then located using energy conservation. Since the EMCCD
efficiency is low in the infrared, it is convenient to restrict the search to
the band where the conjugates of the identified signals are expected: for an
SFWM process the wavelengths obey
\begin{equation}
\frac{1}{\lambda_s}+\frac{1}{\lambda_i}=\frac{2}{\lambda_p},
\label{eq:energy}
\end{equation}
so a pump around $800$~nm places the idlers of the $648$~nm and $660$~nm signals
between $1000$ and $1100$~nm. Scanning this reduced range with longer exposure
times reveals the two idlers at $1021$~nm and $1048$~nm. We have thus identified
two pairs, $648/1048$~nm and $660/1021$~nm, corresponding to a pump at
$800.8$~nm and $801.7$~nm respectively, both consistent with the pump bandwidth.

Having established the spectral properties of the pairs, we now turn to
polarisation: we study how the pump polarisation produces each pair, confirm
through the $g^{(2)}(0)$ measurement that the photons are temporally correlated,
consistent with pair generation, and determine the polarisation of the signal and
idler photons. These properties will be essential for the mode identification of
Section~\ref{sec:modelling}.

\subsection{Pump dependence}
\label{sec:pumpdep}

\begin{figure}[htpb]
\centering
\includegraphics[width=0.8\textwidth]{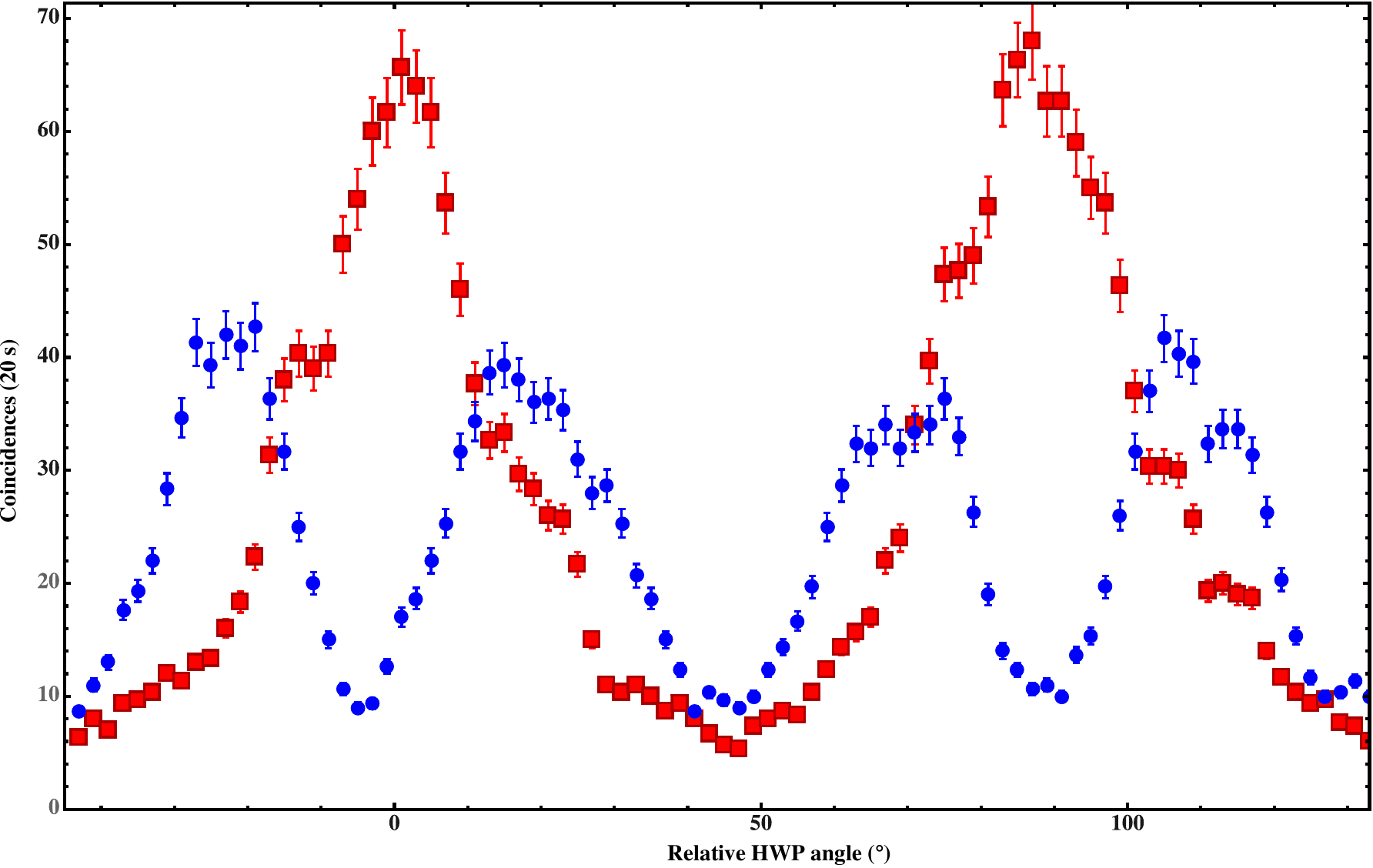}
\caption{Coincidences versus relative half-wave-plate angle for the two pairs.
The $660/1021$~nm pair (blue dots) shows two maxima ($90^\circ$ apart), the
signature of a same-polarisation pump, and the $648/1048$~nm pair (red squares)
shows four maxima ($45^\circ$ apart), the signature of an orthogonal pump. The
zero of the angle axis is not calibrated; only the number and spacing of the
maxima are used.}
\label{fig:pumpdependance}
\end{figure}

To study the influence of the pump polarisation, we record the number of
coincidences between the signal and idler photons as a function of the input
polarisation, in configuration~(ii) of Fig.~\ref{fig:exp_setup}, where the QWP
maximises the contrast of the coincidence modulation and the HWP sweeps the
orientation of the linear polarisation, which decomposes onto the $x$ and $y$
axes of the microcoupler. As shown in Fig.~\ref{fig:pumpdependance}, the
coincidences accumulated over $20$~seconds are plotted against the relative HWP
angle $\theta$, the actual polarisation angle being twice that value,
$\varphi=2\theta$. Over a $180^\circ$ range, the coincidences for the
$660/1021$~nm and $648/1048$~nm photons show two maxima separated by $90^\circ$
and four maxima separated by $45^\circ$ respectively. The wave-plate zero is not
calibrated, but this is irrelevant here, since the diagnostic rests on the
number and spacing of the maxima rather than their absolute position.

This behaviour follows from the SFWM process, which converts two pump photons
into a signal and an idler while conserving energy. Because two pump photons are
consumed, the coincidence rate $R$ scales quadratically with the product of the
pump field components feeding the process~\cite{agrawal}. As the HWP is rotated,
these components are $E_x\sim\cos\varphi$ and $E_y\sim\sin\varphi$. When the two
pump photons share the same polarisation,
\begin{equation}
R\sim(\cos\varphi\,\cos\varphi)^2\sim\cos^4\varphi\ (xx)\quad\text{or}\quad
R\sim\sin^4\varphi\ (yy),
\label{eq:pumpdep_par}
\end{equation}
giving a period of $180^\circ$ in $\varphi$ and two maxima; when they are
orthogonal,
\begin{equation}
R\sim(\cos\varphi\,\sin\varphi)^2\sim\tfrac{1}{4}\sin^2(2\varphi),
\label{eq:pumpdep_perp}
\end{equation}
with a period of $90^\circ$ in $\varphi$ and four maxima.
The four maxima of the $648/1048$~nm pair thus reveal an orthogonal pump,
and the two maxima of the $660/1021$~nm pair a parallel one.

We then verify that the detected photons are temporally correlated, rather
than arising from noise or a Raman signal, by measuring the second-order
cross-correlation $g^{(2)}(0)$ between signal and idler ~\cite{tapster1998}. The pump-dependence
study makes this possible by letting us select the produced pair through the
pump polarisation: having isolated one pair this way, in configuration~(ii) and
with the monochromator providing spectral filtering on the signal arm, we
measure the central-to-side-peak ratio, that is, the ratio of the coincidences in
the zero-delay peak to those in the side peaks, which count accidental coincidences
between photons from different pump pulses. We obtain $g^{(2)}(0)\approx 4$. This
enhanced zero-delay signal--idler coincidence rules out an uncorrelated background
and confirms a temporal correlation consistent with pair generation. Because the
signal and idler autocorrelations were not measured, this result is not used alone
as a Cauchy--Schwarz test of nonclassicality.

From the $g^{(2)}(0)$ measurement and the pump-dependence study, we have thus
shown that the photons of each pair are correlated, and that the $648/1048$~nm
pair is generated by an orthogonal pump and the $660/1021$~nm pair by a parallel
one. Identifying the processes behind each pair, however, also requires the
polarisation of the signal and idler photons themselves.

\subsection{Single-photon polarisation tomography}
\label{sec:tomography}

We reconstruct the polarisation state of each photon of interest by quantum
tomography ~\cite{james2001}, obtaining its density matrix $\rho_m$ and Stokes vector
$\vec{S}_m$ ~\cite{loudon}, two equivalent descriptions of a quantum state related by
$\rho=\tfrac{1}{2}(\mathbb{I}+S_i\sigma_i)$, where the $S_i$ are the
Stokes-vector coordinates and the $\sigma_i$ the Pauli matrices. The
reconstruction uses configuration~(ii): the photon is sent through a
polarisation analyser made of a quarter-wave plate, a half-wave plate and a PBS,
projecting it onto the six states $\lvert H\rangle$, $\lvert V\rangle$, $\lvert D\rangle$, $\lvert A\rangle$,
$\lvert R\rangle$, $\lvert L\rangle$, and the tomography is carried out on coincidences between
the analysed photon and its conjugate, with the monochromator set to the signal
wavelength. The idler is measured by moving the monochromator to the other arm,
and the other pair by readjusting the pump polarisation.

The resulting Stokes vectors of the four photons, with the quantities used to
study their polarisation, are collected in Table~\ref{tab:stokes}. As noted
earlier, propagation through the non-polarisation-maintaining output fibres
applies unknown polarisation transformations $J(\lambda)$, so that the measured
state is $\rho_\mathrm{out}(\lambda)=J(\lambda)\,\rho_\mathrm{waist}\,J^{\dagger}(\lambda)$.
Because signal and idler differ markedly in wavelength they are transformed
differently, and their absolute states cannot be compared. The two signals ($648$
and $660$~nm), and likewise the two idlers ($1021$ and $1048$~nm), differ in
wavelength by less than $3\%$, and we assume that they undergo nearly the same
transformation, $J(\lambda_1)\approx J(\lambda_2)$. This assumption is central to the
analysis, since the distance between two states is preserved only under a common
unitary transformation; it could not be checked by a classical polarimetric
calibration of the output fibres at the four wavelengths, the device having been
cleaved for SEM imaging after the experiments (Sec.~\ref{sec:fabrication}).

Before this comparison, the photons must have a well-defined polarisation state.
The purity $P=\tfrac{1}{2}\!\left(1+\lVert\vec{S}\rVert^2\right)$ of each state,
listed in Table~\ref{tab:purity}, lies between $0.83$ and $0.91$, confirming
states that are fairly close to pure. To quantify whether two photons share a
polarisation, we use the trace distance
\begin{equation}
\mathrm{TD}_{mn}=\tfrac{1}{2}\,\mathrm{Tr}\sqrt{(\rho_m-\rho_n)^2}
=\tfrac{1}{2}\,\mathrm{Tr}|\rho_m-\rho_n|\in[0,1],
\label{eq:td}
\end{equation}
which in terms of Stokes vectors reads
$\tfrac{1}{2}\lVert\vec{S}_m-\vec{S}_n\rVert$, equal to $0$ for identical states
and $1$ for orthogonal pure states~\cite{nielsenchuang}. From
Table~\ref{tab:purity}, the trace distance is $0.35\pm0.03$ between the signals,
whose output polarisations are therefore similar, and $0.84\pm0.05$ between the
idlers, whose output polarisations are nearly orthogonal; the uncertainties follow
from Monte Carlo propagation of the Stokes-vector uncertainties of
Table~\ref{tab:stokes}.

In summary, the $648/1048$~nm pair is generated by orthogonally polarised pump
photons and the $660/1021$~nm pair by pump photons of the same polarisation,
while the two signals have similar output polarisations and the two idlers nearly
orthogonal ones. This information is crucial for determining which processes, among
the six possible modes of the microcoupler, generate the photon pairs by SFWM,
as shown in what follows. Whereas process discrimination in birefringent few-mode fibres has relied on
selectively exciting chosen mode combinations at the input~\cite{majchrowska2022},
here the process is identified after the fact, from the pump-polarisation
dependence and the single-photon tomography together.

\begin{table}[h]
\centering
\caption{Measured single-photon Stokes vectors (lab frame; $S_0=1$;
$1\sigma$ uncertainties).}
\label{tab:stokes}
\begin{tabular}{l c c c}
\hline
Photon & $S_1$ (H/V) & $S_2$ (D/A) & $S_3$ (R/L)\\
\hline
Signal $648$~nm  & $+0.618\pm0.049$ & $-0.324\pm0.055$ & $-0.414\pm0.054$\\
Signal $660$~nm  & $+0.556\pm0.023$ & $-0.621\pm0.023$ & $+0.208\pm0.027$\\
Idler $1021$~nm  & $+0.858\pm0.034$ & $-0.152\pm0.067$ & $+0.257\pm0.068$\\
Idler $1048$~nm  & $-0.710\pm0.085$ & $-0.362\pm0.113$ & $-0.303\pm0.092$\\
\hline
\end{tabular}
\end{table}

\begin{table}[h]
\centering
\caption{Single-photon purities and trace distances between photons of similar
wavelength. The two signals have similar output polarisations, the two idlers nearly
orthogonal ones. Uncertainties are obtained by Monte Carlo propagation of the
Stokes-vector uncertainties of Table~\ref{tab:stokes}.}
\label{tab:purity}
\begin{tabular}{l c c c c}
\hline
 & \multicolumn{2}{c}{signal} & \multicolumn{2}{c}{idler}\\
wavelength (nm) & 648 & 660 & 1021 & 1048\\
\hline
purity $P$ & 0.83 & 0.87 & 0.91 & 0.86\\
trace distance TD & \multicolumn{2}{c}{$0.35\pm0.03$} & \multicolumn{2}{c}{$0.84\pm0.05$}\\
\hline
\end{tabular}
\end{table}

%% file: section4_modelling.tex
\section{Modelling and process identification}
\label{sec:modelling}

The experiment leaves us with two photon pairs and a set of polarisation clues,
but not with their origin: of all the four-wave-mixing processes the microcoupler
can support, which two produced the pairs we detected? This section answers that
question. We first model the elliptical waist and the six modes it guides, then
enumerate every process allowed by symmetry, and finally let the measurements of
Section~\ref{sec:experiment} winnow the list. No single observation settles the
matter alone; phase matching, pump polarisation and single-photon tomography must
be brought to bear together, and their combined observations favour a single pair
of processes within the present geometrical model.

\subsection{Geometry and modes}
\label{sec:geometry}

As shown in figure~\ref{fig:sem_pics}, in the central region where the fibres have
fused the structure can be approximated by an ellipsoidal shape, so modelling the
device requires knowing its two axes. Because the microcoupler is fragile and
the waist position is hard to locate, reading these dimensions directly from a
transverse cross-section at the waist is impractical, and we instead proceed in
steps, assuming the axis ratio stays constant along the central region. The
microcoupler is first cleaved into several sections so that different regions can
be analysed; a section far from the waist is then imaged by SEM to determine the
ellipticity. From this cross-section (figure~\ref{fig:sem_pics}a) the major axis
measures $4.857~\mu$m and the minor axis $3.651~\mu$m, an ellipticity of about
$1.33$ (major/minor), with $x$ the major-axis direction and $y$ the minor. The
region containing the waist is then imaged from the side at several positions to
find its widest diameter, which reaches $1.161~\mu$m (figure~\ref{fig:sem_pics}b).
Combining the ellipticity measured away from the waist with this maximum diameter
gives transverse dimensions at the waist of $1.16~\mu$m by $0.87~\mu$m, consistent with the
$1~\mu$m target set during tapering; the small ellipticity also supports the
assumption of a single fused fibre with an elliptical structure.

\begin{figure}[htbp]
\centering
\begin{tabular}{ccc}
 & $E_{x}$ & $E_{y}$ \\[6pt]
$\mathrm{LP}_{01}^{x}$ &
\includegraphics[scale=0.20]{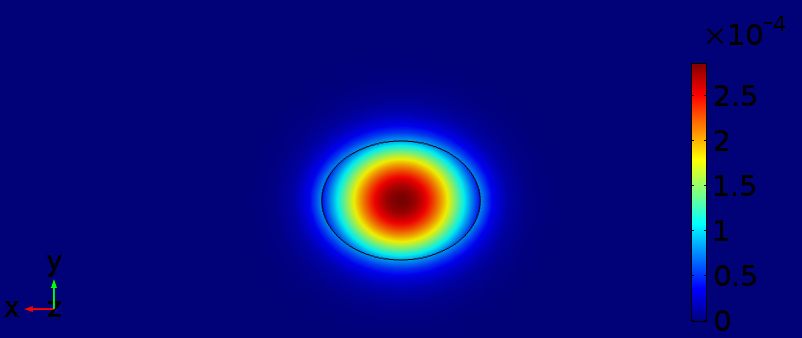} &
\includegraphics[scale=0.20]{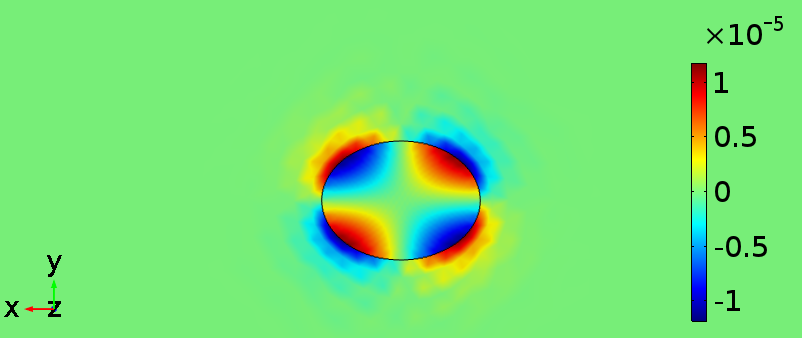} \\[6pt]
$\mathrm{LP}_{01}^{y}$ &
\includegraphics[scale=0.20]{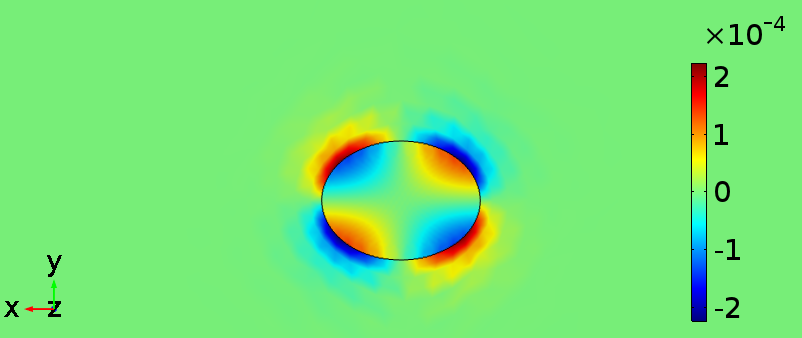} &
\includegraphics[scale=0.20]{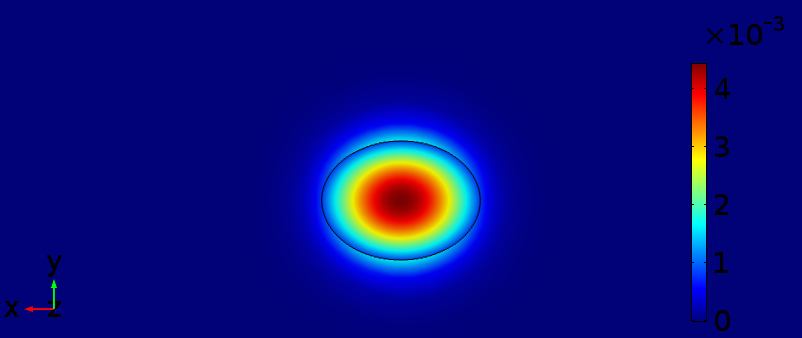} \\[6pt]
$\mathrm{LP}_{11}^{y\text{-even}}$ &
\includegraphics[scale=0.20]{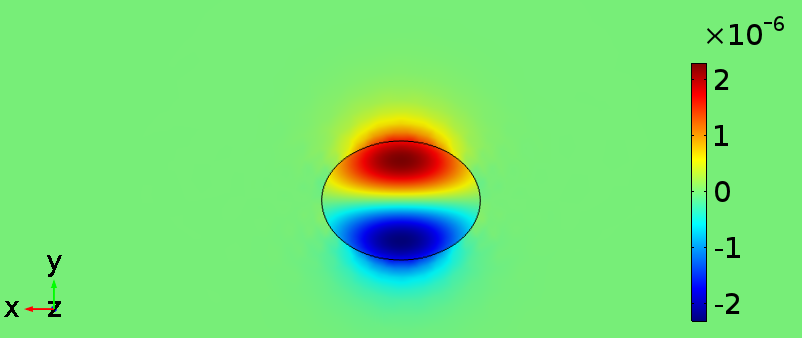} &
\includegraphics[scale=0.20]{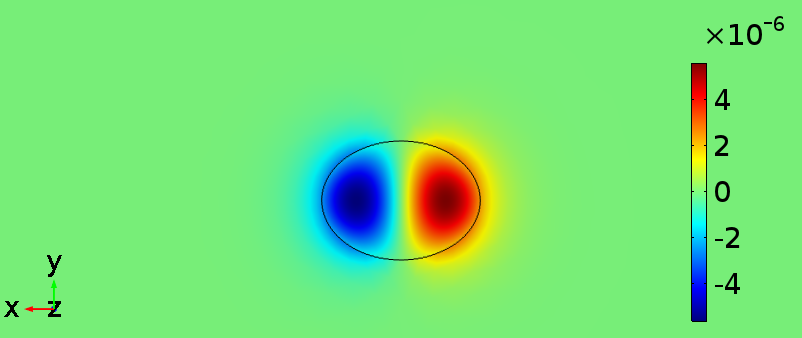} \\[6pt]
$\mathrm{LP}_{11}^{x\text{-even}}$ &
\includegraphics[scale=0.20]{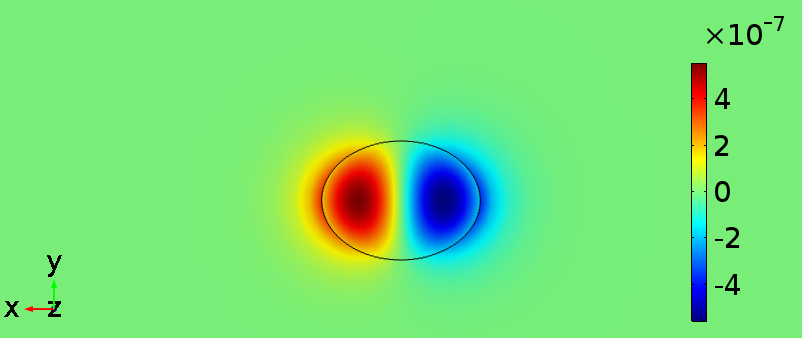} &
\includegraphics[scale=0.20]{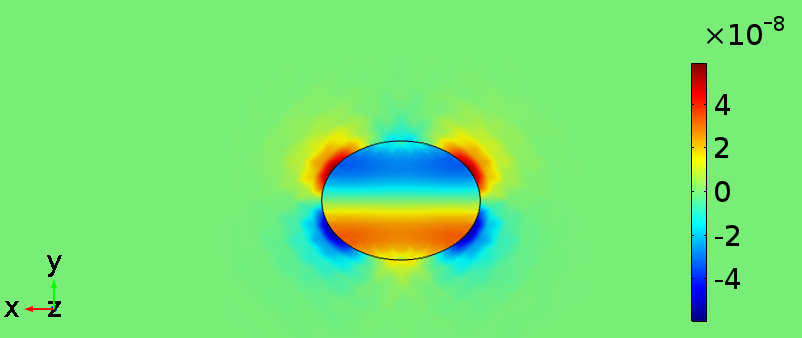} \\[6pt]
$\mathrm{LP}_{11}^{x\text{-odd}}$ &
\includegraphics[scale=0.20]{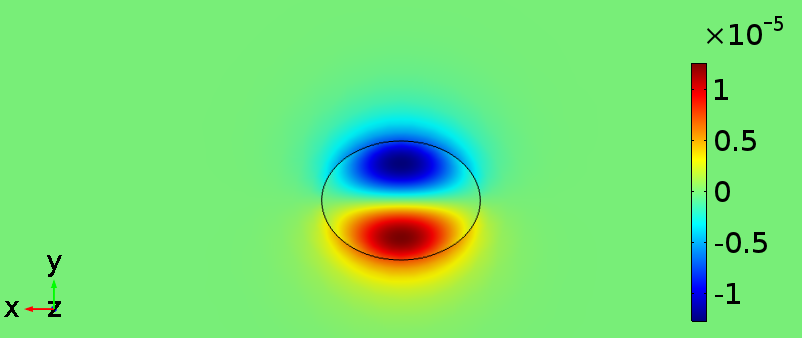} &
\includegraphics[scale=0.20]{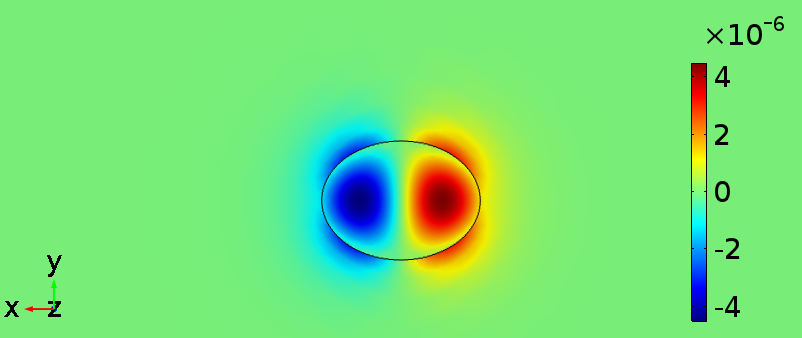} \\[6pt]
$\mathrm{LP}_{11}^{y\text{-odd}}$ &
\includegraphics[scale=0.20]{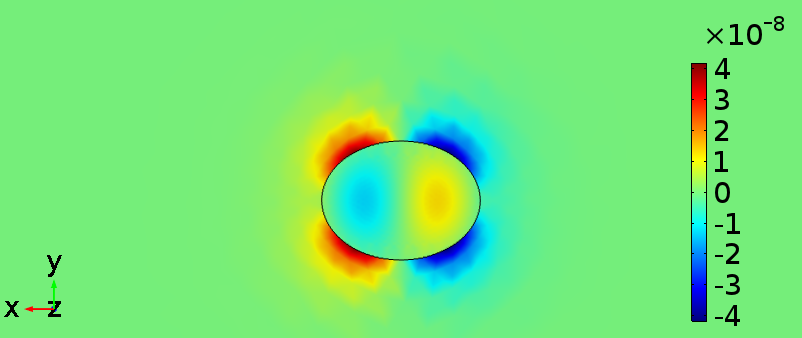} &
\includegraphics[scale=0.20]{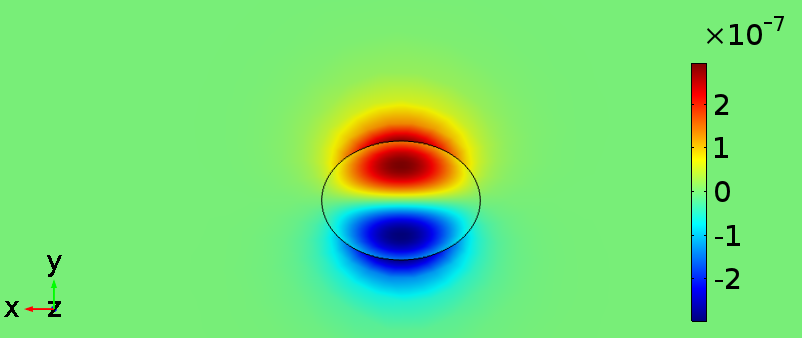} \\
\end{tabular}
\caption{Transverse field profiles of the six guided modes of the microcoupler at
$\lambda=800$~nm: four well-defined modes (the two fundamentals
$\mathrm{LP}_{01}^{x}$, $\mathrm{LP}_{01}^{y}$ and the two
$\mathrm{LP}_{11}^{x\text{-even}}$, $\mathrm{LP}_{11}^{y\text{-odd}}$) and two
hybrid modes ($\mathrm{LP}_{11}^{y\text{-even}}$, $\mathrm{LP}_{11}^{x\text{-odd}}$).
The superscript $x$ or $y$ denotes the polarisation and even/odd the spatial
parity. Modes are shown in the same order as figure~\ref{fig:refrac_index},
by decreasing effective index.}
\label{fig:modes_profiles}
\end{figure}

\begin{figure}[htbp]
\centering
\includegraphics[width=0.6\textwidth]{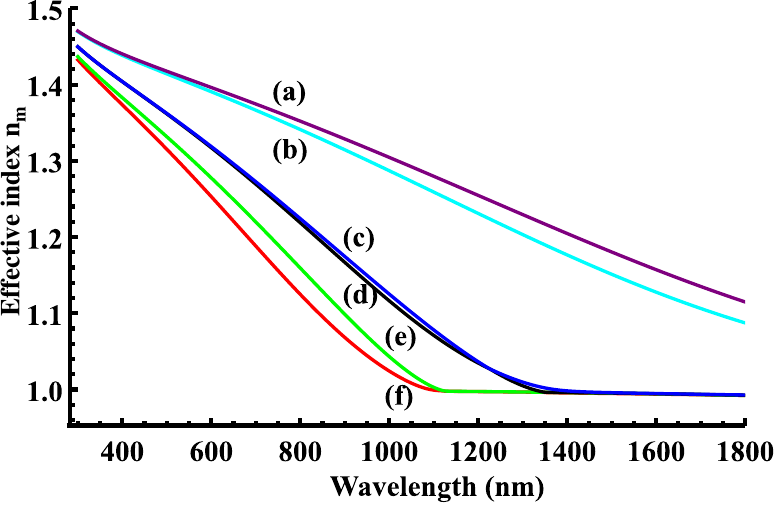}
\caption{Effective indices $n_m$ of the six guided modes versus wavelength over a
$300$ to $1800$~nm sweep, from the COMSOL eigenmode simulations:
(a)~$\mathrm{LP}_{01}^{x}$, (b)~$\mathrm{LP}_{01}^{y}$,
(c)~$\mathrm{LP}_{11}^{y\text{-}\mathrm{even}}$,
(d)~$\mathrm{LP}_{11}^{x\text{-}\mathrm{even}}$,
(e)~$\mathrm{LP}_{11}^{x\text{-}\mathrm{odd}}$,
(f)~$\mathrm{LP}_{11}^{y\text{-}\mathrm{odd}}$. These indices set the propagation
constants $\beta_m=n_m\,\omega/c$ used to solve the phase-matching condition.}
\label{fig:refrac_index}
\end{figure}

With these dimensions known, we use COMSOL, which solves Maxwell's equations on a
user-defined geometry, to find the eigenmodes and their effective indices $n_m$, where $m$ labels the mode.
The device is modelled as fused silica, its index taken from the
Sellmeier equations, surrounded by a medium of index $n=1$, on a non-uniform grid
that is fine near the centre where the field is most concentrated and coarser far
from the device. The simulations show six supported modes
(figure~\ref{fig:modes_profiles}), and a wavelength sweep gives their effective
indices (figure~\ref{fig:refrac_index}) at several wavelengths. For convenience we
label them with the usual LP nomenclature of a weakly guiding
fibre~\cite{gloge1971}, keeping in mind that the LP modes are not the exact
eigenmodes but vector modes~\cite{kogelnikwinzer2012}. From the simulated field
components (figure~\ref{fig:modes_profiles}) we sort them into two groups. Four are
well defined: the two fundamental modes with orthogonal $x$ and $y$ polarisations
and a single lobe (LP$_{01}^{x}$, LP$_{01}^{y}$), and the two LP$_{11}$ modes with
orthogonal polarisations whose two-lobe structure is set by their nodal plane,
LP$_{11}^{x\text{-even}}$ with a vertical nodal plane and LP$_{11}^{y\text{-odd}}$
with a horizontal one. The remaining two, which we call hybrid, show a two-lobe
structure in both $E_x$ and $E_y$; as these components differ in amplitude, each is
named after the stronger one: LP$_{11}^{y\text{-even}}$ (vertical nodal plane),
whose $E_y$ is about twice its $E_x$, and LP$_{11}^{x\text{-odd}}$ (horizontal nodal
plane), whose $E_x$ is about $2.5$ times its $E_y$.

In principle an elliptical core lifts the LP$_{11}$ degeneracy completely,
splitting it into two modes of well-defined odd/even symmetry~\cite{wang2005}, but
a clean separation requires a strong ellipticity, typically an axis ratio of order
$2$ or more in low-index-contrast fibres. The axis ratio here is well below that
($1.33$), so an incomplete resolution is unsurprising: the two fundamentals
together with LP$_{11}^{x\text{-even}}$ and LP$_{11}^{y\text{-odd}}$ stay well
defined, while the COMSOL simulations show the other two retaining a hybrid
structure that mixes $E_x$ and $E_y$~\cite{kogelnikwinzer2012}. These simulations
thus provide the six modes and their effective indices, which we now use to
identify the nonlinear processes at work in the microcoupler.

\subsection{Selection rules and counting}
\label{sec:selectionrules}

SFWM is a third-order nonlinear process that annihilates two pump photons and
creates a signal and an idler obeying the phase-matching condition
\begin{equation}
\Delta k=\beta_{p_1}(\omega_p)+\beta_{p_2}(\omega_p)-\beta_s(\omega_s)-\beta_i(\omega_i)=0,
\label{eq:deltak}
\end{equation}
where $\beta_m=n_m\,\omega/c$ is the propagation constant of mode $m$ and $n_m$ its
effective index. This condition retains only the linear propagation constants: the
nonlinear contribution $\Delta k_\mathrm{NL}$ from self- and cross-phase modulation of the
pump is neglected. For the highest average pump power used in this work, about $20$~mW
(Section~\ref{sec:source}), i.e.\ about $0.25$~nJ per pulse and a peak power of about
$2.5$~kW, and a nonlinear coefficient $\gamma\approx0.26$~W$^{-1}$m$^{-1}$ estimated from
the waist dimensions (Section~\ref{sec:geometry}) with
$n_2=2.6\times10^{-20}$~m$^2$/W~\cite{agrawal}, this contribution is bounded by
$\Delta k_\mathrm{NL}\approx2\gamma P\approx1.3\times10^{3}$~m$^{-1}$, an upper value since
the pump is shared among several modes and broadens by dispersion.
The pair must also conserve energy~(\ref{eq:energy}); given the
form of~(\ref{eq:deltak}), it is convenient to work in frequency and to write energy
conservation as $\omega_s+\omega_i=2\omega_p$.
The process is governed by the third-order susceptibility tensor $\chi^{(3)}$. Because
the signal and idler start from vacuum, spontaneous pair generation cannot be described
by classical coupled-amplitude equations and requires quantised signal and idler
fields. Following the standard treatment of SFWM~\cite{garaypalmett2007}, a process
$\mu$ in which pump photons in modes $p_1$ and $p_2$ generate a signal in mode $s$ and
an idler in mode $i$ produces pairs with the joint spectral amplitude
\begin{equation}
\mathcal{F}_\mu(\omega_s,\omega_i)\propto\alpha(\omega_s+\omega_i)\,\Gamma_\mu\,\Phi_\mu(\omega_s,\omega_i),
\label{eq:jsa}
\end{equation}
where $\alpha$ is the spectral envelope of the pump,
\begin{equation}
\Phi_\mu(\omega_s,\omega_i)=\int_0^L\mathrm{d}z\,
\exp\!\left(i\int_0^{z}\Delta k_\mu(z')\,\mathrm{d}z'\right)
\label{eq:phasefunction}
\end{equation}
is the phase-matching function, which reduces to
$L\,\mathrm{sinc}(\Delta k_\mu L/2)\,e^{i\Delta k_\mu L/2}$ for a guide of uniform section
and length $L$ and peaks at $\Delta k_\mu=0$, and
\begin{equation}
\Gamma_\mu=\int\mathrm{d}A\;\chi^{(3)}_{abcd}\,F_{s,c}^{*}F_{i,d}^{*}F_{p_1,a}F_{p_2,b}
\label{eq:gamma}
\end{equation}
is the vectorial nonlinear overlap, $F_{j,a}(x,y)$ being the Cartesian component $a$ of
the transverse profile of mode $j$, with summation over repeated indices. For modes of
well-defined polarisation, $\Gamma_\mu$ factorises as $\chi_\mathrm{eff}\,f_{s\,i\,p_1p_2}$,
where $\chi_\mathrm{eff}$ is the effective component of $\chi^{(3)}$ for the chosen
polarisation combination, with $F_j$ the transverse profile of the dominant field
component of mode $j$, and
\begin{equation}
f_{ijkl}=\frac{\langle F_i^{*}F_j^{*}F_k F_l\rangle}
{\left(\langle|F_i|^2\rangle\langle|F_j|^2\rangle\langle|F_k|^2\rangle\langle|F_l|^2\rangle\right)^{1/2}}
\label{eq:overlap}
\end{equation}
is the overlap integral between the four modes involved.

Each SFWM process obeys two selection rules, a spatial one tied to the overlap
integral and a polarisation one tied to $\chi_\mathrm{eff}$, and is forbidden if
either vanishes~\cite{garaypalmett2016}. The overlap integral is nonzero, and the combination allowed, when
the product of the four mode profiles $F_i F_j F_k F_l$ is even; an odd product
makes it vanish by symmetry. This parity is carried by the LP$_{11}$ modes, whose
two lobes have opposite signs, while the single-lobed LP$_{01}$ modes leave it
unchanged. On the polarisation side, since fused silica is
isotropic~\cite{boyd,agrawal} the susceptibility reduces to
\begin{equation}
\chi^{(3)}_{ijkl}=\chi_{XXYY}\,\delta_{ij}\delta_{kl}+\chi_{XYXY}\,\delta_{ik}\delta_{jl}
+\chi_{XYYX}\,\delta_{il}\delta_{jk}.
\label{eq:chidelta}
\end{equation}
The Kronecker deltas are nonzero only for matching indices, so the surviving
components are those in which $X$ and $Y$ each appear an even number of times, eight
in all; isotropy further makes the $\chi_{ijkl}$ invariant under exchange of $X$ and
$Y$, pairing them as $\chi_{XXXX}=\chi_{YYYY}$, $\chi_{XYXY}=\chi_{YXYX}$,
$\chi_{XYYX}=\chi_{YXXY}$ and $\chi_{XXYY}=\chi_{YYXX}$, and leaving four values
related through equation~(\ref{eq:chidelta}) by
$\chi_{XXXX}=\chi_{XXYY}+\chi_{XYXY}+\chi_{XYYX}$; for the electronic response of
silica the three off-diagonal components are equal, each being
$\chi_{XXXX}/3$~\cite{boyd}.

For each process, four modes, two for the pump and one each for signal and idler,
are drawn from the six. The pump is injected in the fundamental mode of the input
fibre, which is no longer an eigenmode of the fused region and there spreads over
all six modes, so each pump photon can occupy any of them, and several of the four
photons may share a mode, for instance a pump photon and the signal. Counting the
distinct processes then amounts to removing the combinations forbidden by the two
selection rules and discarding those that differ only by relabelling, such as two
pump photons of the same wavelength. After these filters, 51 allowed and
independent SFWM processes remain, listed in table~\ref{tab:combinations}. These
rules give the physically possible processes but not which of the 51 actually
occur. The same intermodal-vectorial processes have been exploited in birefringent
few-mode fibres to generate and identify multiple signal-idler bands by selective
modal excitation~\cite{majchrowska2022} and to propose entanglement across
several degrees of freedom~\cite{gawlik2025}.

\definecolor{cand}{gray}{0.85}
\setlength{\LTcapwidth}{\textwidth}
{\small
\begin{longtable}{cl c llcc}
\caption{The 51 SFWM combinations allowed by the parity and $\chi^{(3)}$
selection rules, with the phase-matching solutions for a pump near 800~nm.
Columns give the combination number, the two pump modes, the two daughter
modes, the polarisation structure (pump then daughters, read from the modes),
and the shorter ($\lambda_{\min}$, signal) and longer ($\lambda_{\max}$, idler)
phase-matched wavelengths. A dash means $\Delta k=0$ has no solution near the
pump. Combinations 23 and 49 each admit several solutions and are listed on
successive lines. Highlighted rows are the four candidates consistent with a
detected pair (11, 21, 32, 51). A star on the number marks the combinations
reordered from notebook order to place the signal ($\lambda_{\min}$) first
(21, 27, 51). The two hybrid modes are $\mathrm{LP}_{11}^{x\text{-}\mathrm{odd}}$ and $\mathrm{LP}_{11}^{y\text{-}\mathrm{even}}$ (see text).}
\label{tab:combinations}\\
\hline
\# & pump & & daughters & pol & $\lambda_{\min}$ (nm) & $\lambda_{\max}$ (nm)\\
\hline
\endfirsthead
\multicolumn{7}{c}{\tablename~\thetable{} (continued)}\\
\hline
\# & pump & & daughters & pol & $\lambda_{\min}$ (nm) & $\lambda_{\max}$ (nm)\\
\hline
\endhead
\hline
\endfoot
1 & $\mathrm{LP}_{11}^{y\text{-}\mathrm{odd}}+\mathrm{LP}_{11}^{y\text{-}\mathrm{odd}}$ & $\rightarrow$ & $\mathrm{LP}_{11}^{y\text{-}\mathrm{odd}}+\mathrm{LP}_{11}^{y\text{-}\mathrm{odd}}$ & $yy\!\to\!y+y$ & -- & -- \\
2 & $\mathrm{LP}_{11}^{y\text{-}\mathrm{odd}}+\mathrm{LP}_{11}^{y\text{-}\mathrm{odd}}$ & $\rightarrow$ & $\mathrm{LP}_{11}^{x\text{-}\mathrm{even}}+\mathrm{LP}_{11}^{x\text{-}\mathrm{even}}$ & $yy\!\to\!x+x$ & -- & -- \\
3 & $\mathrm{LP}_{11}^{x\text{-}\mathrm{even}}+\mathrm{LP}_{11}^{x\text{-}\mathrm{even}}$ & $\rightarrow$ & $\mathrm{LP}_{11}^{x\text{-}\mathrm{even}}+\mathrm{LP}_{11}^{x\text{-}\mathrm{even}}$ & $xx\!\to\!x+x$ & 620.7 & 1125.0 \\
4 & $\mathrm{LP}_{11}^{x\text{-}\mathrm{even}}+\mathrm{LP}_{11}^{x\text{-}\mathrm{even}}$ & $\rightarrow$ & $\mathrm{LP}_{11}^{y\text{-}\mathrm{odd}}+\mathrm{LP}_{11}^{y\text{-}\mathrm{odd}}$ & $xx\!\to\!y+y$ & -- & -- \\
5 & $\mathrm{LP}_{11}^{y\text{-}\mathrm{odd}}+\mathrm{LP}_{11}^{x\text{-}\mathrm{even}}$ & $\rightarrow$ & $\mathrm{LP}_{11}^{x\text{-}\mathrm{even}}+\mathrm{LP}_{11}^{y\text{-}\mathrm{odd}}$ & $xy\!\to\!x+y$ & 796.0 & 804.0 \\
6 & $\mathrm{LP}_{11}^{y\text{-}\mathrm{odd}}+\mathrm{LP}_{11}^{y\text{-}\mathrm{odd}}$ & $\rightarrow$ & $\mathrm{LP}_{11}^{x\text{-}\mathrm{odd}}+\mathrm{LP}_{11}^{x\text{-}\mathrm{odd}}$ & $yy\!\to\!x+x$ & -- & -- \\
7 & $\mathrm{LP}_{11}^{y\text{-}\mathrm{odd}}+\mathrm{LP}_{11}^{y\text{-}\mathrm{odd}}$ & $\rightarrow$ & $\mathrm{LP}_{11}^{y\text{-}\mathrm{even}}+\mathrm{LP}_{11}^{y\text{-}\mathrm{even}}$ & $yy\!\to\!y+y$ & -- & -- \\
8 & $\mathrm{LP}_{11}^{x\text{-}\mathrm{even}}+\mathrm{LP}_{11}^{x\text{-}\mathrm{even}}$ & $\rightarrow$ & $\mathrm{LP}_{11}^{y\text{-}\mathrm{even}}+\mathrm{LP}_{11}^{y\text{-}\mathrm{even}}$ & $xx\!\to\!y+y$ & -- & -- \\
9 & $\mathrm{LP}_{11}^{x\text{-}\mathrm{even}}+\mathrm{LP}_{11}^{x\text{-}\mathrm{even}}$ & $\rightarrow$ & $\mathrm{LP}_{11}^{x\text{-}\mathrm{odd}}+\mathrm{LP}_{11}^{x\text{-}\mathrm{odd}}$ & $xx\!\to\!x+x$ & -- & -- \\
10 & $\mathrm{LP}_{11}^{y\text{-}\mathrm{odd}}+\mathrm{LP}_{11}^{x\text{-}\mathrm{even}}$ & $\rightarrow$ & $\mathrm{LP}_{11}^{y\text{-}\mathrm{even}}+\mathrm{LP}_{11}^{x\text{-}\mathrm{odd}}$ & $xy\!\to\!y+x$ & -- & -- \\
\rowcolor{cand} \textbf{11} & $\mathrm{LP}_{11}^{x\text{-}\mathrm{odd}}+\mathrm{LP}_{11}^{x\text{-}\mathrm{odd}}$ & $\rightarrow$ & $\mathrm{LP}_{11}^{x\text{-}\mathrm{odd}}+\mathrm{LP}_{11}^{x\text{-}\mathrm{odd}}$ & $xx\!\to\!x+x$ & 647.6 & 1046.3 \\
12 & $\mathrm{LP}_{11}^{x\text{-}\mathrm{odd}}+\mathrm{LP}_{11}^{x\text{-}\mathrm{odd}}$ & $\rightarrow$ & $\mathrm{LP}_{11}^{y\text{-}\mathrm{even}}+\mathrm{LP}_{11}^{y\text{-}\mathrm{even}}$ & $xx\!\to\!y+y$ & -- & -- \\
13 & $\mathrm{LP}_{11}^{y\text{-}\mathrm{even}}+\mathrm{LP}_{11}^{y\text{-}\mathrm{even}}$ & $\rightarrow$ & $\mathrm{LP}_{11}^{y\text{-}\mathrm{even}}+\mathrm{LP}_{11}^{y\text{-}\mathrm{even}}$ & $yy\!\to\!y+y$ & -- & -- \\
14 & $\mathrm{LP}_{11}^{y\text{-}\mathrm{even}}+\mathrm{LP}_{11}^{y\text{-}\mathrm{even}}$ & $\rightarrow$ & $\mathrm{LP}_{11}^{x\text{-}\mathrm{odd}}+\mathrm{LP}_{11}^{x\text{-}\mathrm{odd}}$ & $yy\!\to\!x+x$ & -- & -- \\
15 & $\mathrm{LP}_{11}^{x\text{-}\mathrm{odd}}+\mathrm{LP}_{11}^{y\text{-}\mathrm{even}}$ & $\rightarrow$ & $\mathrm{LP}_{11}^{y\text{-}\mathrm{even}}+\mathrm{LP}_{11}^{x\text{-}\mathrm{odd}}$ & $xy\!\to\!y+x$ & 616.1 & 1140.5 \\
16 & $\mathrm{LP}_{11}^{x\text{-}\mathrm{odd}}+\mathrm{LP}_{11}^{x\text{-}\mathrm{odd}}$ & $\rightarrow$ & $\mathrm{LP}_{11}^{y\text{-}\mathrm{odd}}+\mathrm{LP}_{11}^{y\text{-}\mathrm{odd}}$ & $xx\!\to\!y+y$ & -- & -- \\
17 & $\mathrm{LP}_{11}^{x\text{-}\mathrm{odd}}+\mathrm{LP}_{11}^{x\text{-}\mathrm{odd}}$ & $\rightarrow$ & $\mathrm{LP}_{11}^{x\text{-}\mathrm{even}}+\mathrm{LP}_{11}^{x\text{-}\mathrm{even}}$ & $xx\!\to\!x+x$ & -- & -- \\
18 & $\mathrm{LP}_{11}^{y\text{-}\mathrm{even}}+\mathrm{LP}_{11}^{y\text{-}\mathrm{even}}$ & $\rightarrow$ & $\mathrm{LP}_{11}^{x\text{-}\mathrm{even}}+\mathrm{LP}_{11}^{x\text{-}\mathrm{even}}$ & $yy\!\to\!x+x$ & -- & -- \\
19 & $\mathrm{LP}_{11}^{y\text{-}\mathrm{even}}+\mathrm{LP}_{11}^{y\text{-}\mathrm{even}}$ & $\rightarrow$ & $\mathrm{LP}_{11}^{y\text{-}\mathrm{odd}}+\mathrm{LP}_{11}^{y\text{-}\mathrm{odd}}$ & $yy\!\to\!y+y$ & -- & -- \\
20 & $\mathrm{LP}_{11}^{x\text{-}\mathrm{odd}}+\mathrm{LP}_{11}^{y\text{-}\mathrm{even}}$ & $\rightarrow$ & $\mathrm{LP}_{11}^{x\text{-}\mathrm{even}}+\mathrm{LP}_{11}^{y\text{-}\mathrm{odd}}$ & $xy\!\to\!x+y$ & 607.7 & 1170.4 \\
\rowcolor{cand} \textbf{21$^{*}$} & $\mathrm{LP}_{11}^{y\text{-}\mathrm{odd}}+\mathrm{LP}_{11}^{x\text{-}\mathrm{odd}}$ & $\rightarrow$ & $\mathrm{LP}_{11}^{y\text{-}\mathrm{odd}}+\mathrm{LP}_{11}^{x\text{-}\mathrm{odd}}$ & $xy\!\to\!y+x$ & 658.0 & 1020.1 \\
22 & $\mathrm{LP}_{11}^{y\text{-}\mathrm{odd}}+\mathrm{LP}_{11}^{x\text{-}\mathrm{odd}}$ & $\rightarrow$ & $\mathrm{LP}_{11}^{y\text{-}\mathrm{even}}+\mathrm{LP}_{11}^{x\text{-}\mathrm{even}}$ & $xy\!\to\!y+x$ & -- & -- \\
23 & $\mathrm{LP}_{11}^{x\text{-}\mathrm{even}}+\mathrm{LP}_{11}^{y\text{-}\mathrm{even}}$ & $\rightarrow$ & $\mathrm{LP}_{11}^{y\text{-}\mathrm{even}}+\mathrm{LP}_{11}^{x\text{-}\mathrm{even}}$ & $xy\!\to\!y+x$ & 601.4 & 1194.6 \\
 & & & & & 626.8 & 1105.6 \\
 & & & & & 653.7 & 1030.6 \\
24 & $\mathrm{LP}_{11}^{x\text{-}\mathrm{even}}+\mathrm{LP}_{11}^{y\text{-}\mathrm{even}}$ & $\rightarrow$ & $\mathrm{LP}_{11}^{x\text{-}\mathrm{odd}}+\mathrm{LP}_{11}^{y\text{-}\mathrm{odd}}$ & $xy\!\to\!x+y$ & -- & -- \\
25 & $\mathrm{LP}_{11}^{y\text{-}\mathrm{odd}}+\mathrm{LP}_{11}^{y\text{-}\mathrm{even}}$ & $\rightarrow$ & $\mathrm{LP}_{11}^{y\text{-}\mathrm{even}}+\mathrm{LP}_{11}^{y\text{-}\mathrm{odd}}$ & $yy\!\to\!y+y$ & 708.3 & 918.9 \\
26 & $\mathrm{LP}_{11}^{y\text{-}\mathrm{odd}}+\mathrm{LP}_{11}^{y\text{-}\mathrm{even}}$ & $\rightarrow$ & $\mathrm{LP}_{11}^{x\text{-}\mathrm{odd}}+\mathrm{LP}_{11}^{x\text{-}\mathrm{even}}$ & $yy\!\to\!x+x$ & -- & -- \\
27$^{*}$ & $\mathrm{LP}_{11}^{x\text{-}\mathrm{even}}+\mathrm{LP}_{11}^{x\text{-}\mathrm{odd}}$ & $\rightarrow$ & $\mathrm{LP}_{11}^{x\text{-}\mathrm{even}}+\mathrm{LP}_{11}^{x\text{-}\mathrm{odd}}$ & $xx\!\to\!x+x$ & 626.5 & 1106.5 \\
28 & $\mathrm{LP}_{11}^{x\text{-}\mathrm{even}}+\mathrm{LP}_{11}^{x\text{-}\mathrm{odd}}$ & $\rightarrow$ & $\mathrm{LP}_{11}^{y\text{-}\mathrm{even}}+\mathrm{LP}_{11}^{y\text{-}\mathrm{odd}}$ & $xx\!\to\!y+y$ & 625.5 & 1109.5 \\
29 & $\mathrm{LP}_{11}^{y\text{-}\mathrm{odd}}+\mathrm{LP}_{11}^{y\text{-}\mathrm{odd}}$ & $\rightarrow$ & $\mathrm{LP}_{01}^{x}+\mathrm{LP}_{01}^{x}$ & $yy\!\to\!x+x$ & -- & -- \\
30 & $\mathrm{LP}_{01}^{x}+\mathrm{LP}_{01}^{x}$ & $\rightarrow$ & $\mathrm{LP}_{01}^{x}+\mathrm{LP}_{01}^{x}$ & $xx\!\to\!x+x$ & -- & -- \\
31 & $\mathrm{LP}_{01}^{x}+\mathrm{LP}_{01}^{x}$ & $\rightarrow$ & $\mathrm{LP}_{11}^{y\text{-}\mathrm{odd}}+\mathrm{LP}_{11}^{y\text{-}\mathrm{odd}}$ & $xx\!\to\!y+y$ & -- & -- \\
\rowcolor{cand} \textbf{32} & $\mathrm{LP}_{11}^{y\text{-}\mathrm{odd}}+\mathrm{LP}_{01}^{x}$ & $\rightarrow$ & $\mathrm{LP}_{01}^{x}+\mathrm{LP}_{11}^{y\text{-}\mathrm{odd}}$ & $xy\!\to\!x+y$ & 630.2 & 1095.2 \\
33 & $\mathrm{LP}_{11}^{y\text{-}\mathrm{odd}}+\mathrm{LP}_{11}^{y\text{-}\mathrm{odd}}$ & $\rightarrow$ & $\mathrm{LP}_{01}^{y}+\mathrm{LP}_{01}^{y}$ & $yy\!\to\!y+y$ & -- & -- \\
34 & $\mathrm{LP}_{01}^{x}+\mathrm{LP}_{01}^{x}$ & $\rightarrow$ & $\mathrm{LP}_{01}^{y}+\mathrm{LP}_{01}^{y}$ & $xx\!\to\!y+y$ & -- & -- \\
35 & $\mathrm{LP}_{01}^{x}+\mathrm{LP}_{01}^{x}$ & $\rightarrow$ & $\mathrm{LP}_{11}^{x\text{-}\mathrm{odd}}+\mathrm{LP}_{11}^{x\text{-}\mathrm{odd}}$ & $xx\!\to\!x+x$ & -- & -- \\
36 & $\mathrm{LP}_{11}^{y\text{-}\mathrm{odd}}+\mathrm{LP}_{01}^{x}$ & $\rightarrow$ & $\mathrm{LP}_{01}^{y}+\mathrm{LP}_{11}^{x\text{-}\mathrm{odd}}$ & $xy\!\to\!y+x$ & -- & -- \\
37 & $\mathrm{LP}_{11}^{x\text{-}\mathrm{odd}}+\mathrm{LP}_{11}^{x\text{-}\mathrm{odd}}$ & $\rightarrow$ & $\mathrm{LP}_{01}^{y}+\mathrm{LP}_{01}^{y}$ & $xx\!\to\!y+y$ & -- & -- \\
38 & $\mathrm{LP}_{01}^{y}+\mathrm{LP}_{01}^{y}$ & $\rightarrow$ & $\mathrm{LP}_{01}^{y}+\mathrm{LP}_{01}^{y}$ & $yy\!\to\!y+y$ & -- & -- \\
39 & $\mathrm{LP}_{01}^{y}+\mathrm{LP}_{01}^{y}$ & $\rightarrow$ & $\mathrm{LP}_{11}^{x\text{-}\mathrm{odd}}+\mathrm{LP}_{11}^{x\text{-}\mathrm{odd}}$ & $yy\!\to\!x+x$ & -- & -- \\
40 & $\mathrm{LP}_{11}^{x\text{-}\mathrm{odd}}+\mathrm{LP}_{01}^{y}$ & $\rightarrow$ & $\mathrm{LP}_{01}^{y}+\mathrm{LP}_{11}^{x\text{-}\mathrm{odd}}$ & $xy\!\to\!y+x$ & -- & -- \\
41 & $\mathrm{LP}_{11}^{x\text{-}\mathrm{odd}}+\mathrm{LP}_{11}^{x\text{-}\mathrm{odd}}$ & $\rightarrow$ & $\mathrm{LP}_{01}^{x}+\mathrm{LP}_{01}^{x}$ & $xx\!\to\!x+x$ & -- & -- \\
42 & $\mathrm{LP}_{01}^{y}+\mathrm{LP}_{01}^{y}$ & $\rightarrow$ & $\mathrm{LP}_{01}^{x}+\mathrm{LP}_{01}^{x}$ & $yy\!\to\!x+x$ & -- & -- \\
43 & $\mathrm{LP}_{01}^{y}+\mathrm{LP}_{01}^{y}$ & $\rightarrow$ & $\mathrm{LP}_{11}^{y\text{-}\mathrm{odd}}+\mathrm{LP}_{11}^{y\text{-}\mathrm{odd}}$ & $yy\!\to\!y+y$ & -- & -- \\
44 & $\mathrm{LP}_{11}^{x\text{-}\mathrm{odd}}+\mathrm{LP}_{01}^{y}$ & $\rightarrow$ & $\mathrm{LP}_{01}^{x}+\mathrm{LP}_{11}^{y\text{-}\mathrm{odd}}$ & $xy\!\to\!x+y$ & -- & -- \\
45 & $\mathrm{LP}_{11}^{y\text{-}\mathrm{odd}}+\mathrm{LP}_{11}^{x\text{-}\mathrm{odd}}$ & $\rightarrow$ & $\mathrm{LP}_{01}^{y}+\mathrm{LP}_{01}^{x}$ & $xy\!\to\!y+x$ & -- & -- \\
46 & $\mathrm{LP}_{01}^{x}+\mathrm{LP}_{01}^{y}$ & $\rightarrow$ & $\mathrm{LP}_{01}^{y}+\mathrm{LP}_{01}^{x}$ & $xy\!\to\!y+x$ & 693.5 & 945.2 \\
47 & $\mathrm{LP}_{01}^{x}+\mathrm{LP}_{01}^{y}$ & $\rightarrow$ & $\mathrm{LP}_{11}^{x\text{-}\mathrm{odd}}+\mathrm{LP}_{11}^{y\text{-}\mathrm{odd}}$ & $xy\!\to\!x+y$ & -- & -- \\
48 & $\mathrm{LP}_{11}^{y\text{-}\mathrm{odd}}+\mathrm{LP}_{01}^{y}$ & $\rightarrow$ & $\mathrm{LP}_{01}^{y}+\mathrm{LP}_{11}^{y\text{-}\mathrm{odd}}$ & $yy\!\to\!y+y$ & 625.0 & 1111.3 \\
49 & $\mathrm{LP}_{11}^{y\text{-}\mathrm{odd}}+\mathrm{LP}_{01}^{y}$ & $\rightarrow$ & $\mathrm{LP}_{11}^{x\text{-}\mathrm{odd}}+\mathrm{LP}_{01}^{x}$ & $yy\!\to\!x+x$ & 625.4 & 1109.7 \\
 & & & & & 664.0 & 1006.1 \\
50 & $\mathrm{LP}_{01}^{x}+\mathrm{LP}_{11}^{x\text{-}\mathrm{odd}}$ & $\rightarrow$ & $\mathrm{LP}_{11}^{x\text{-}\mathrm{odd}}+\mathrm{LP}_{01}^{x}$ & $xx\!\to\!x+x$ & -- & -- \\
\rowcolor{cand} \textbf{51$^{*}$} & $\mathrm{LP}_{01}^{x}+\mathrm{LP}_{11}^{x\text{-}\mathrm{odd}}$ & $\rightarrow$ & $\mathrm{LP}_{11}^{y\text{-}\mathrm{odd}}+\mathrm{LP}_{01}^{y}$ & $xx\!\to\!y+y$ & 656.5 & 1023.8 \\
\end{longtable}
}

\subsection{From allowed processes to compatible candidates}
\label{sec:candidates}

\begin{figure}[htbp]
\centering
\includegraphics[width=\textwidth]{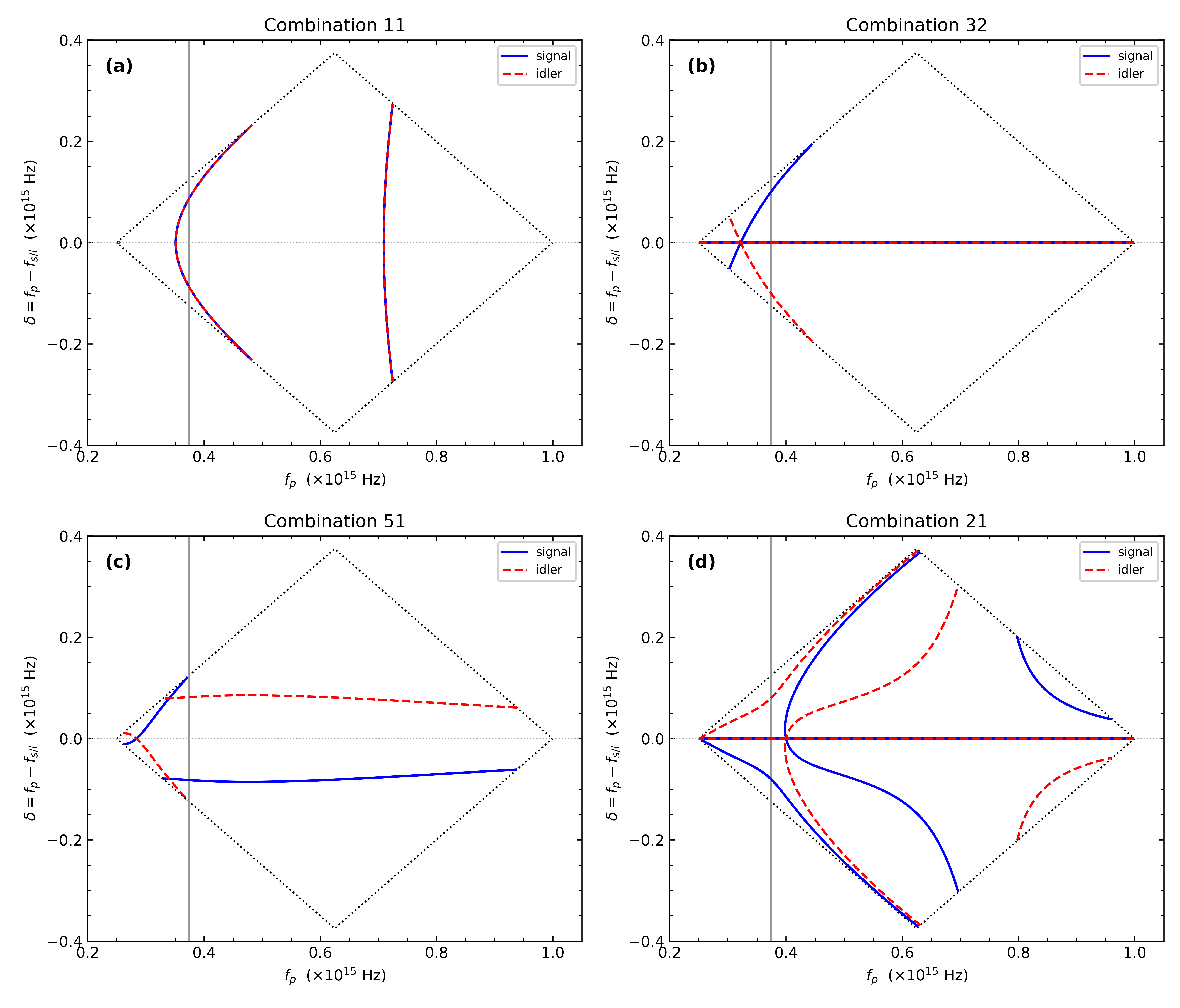}
\caption{Phase-matching curves ($\Delta k=0$) in the $(f_p,\delta)$ plane for the
four candidate combinations: (a)~11, (b)~32, (c)~51 and (d)~21. Solid blue marks
the signal branch and dashed red the idler branch; the dotted diamond is the
energy-conservation boundary. The vertical grey line is the experimental pump
($\lambda_p=800$~nm, i.e.\ $f_p=0.375\times10^{15}$~Hz), and its intersections
with the branches give the phase-matched signal and idler wavelengths reported in
table~\ref{tab:combinations}. For each panel, the offset $|\delta|$ and the
corresponding wavelengths $(\lambda_s,\lambda_i)$ are: (a)~11,
$|\delta|=0.088\times10^{15}$~Hz, $(647.6,\,1046.3)$~nm; (b)~32,
$|\delta|=0.101\times10^{15}$~Hz, $(630.2,\,1095.2)$~nm; (c)~51,
$|\delta|=0.082\times10^{15}$~Hz, $(656.5,\,1023.8)$~nm; (d)~21,
$|\delta|=0.081\times10^{15}$~Hz, $(658.0,\,1020.1)$~nm.}
\label{fig:phasematching_grid}
\end{figure}

Not all 51 allowed processes can generate pairs in our configuration: this requires
the phase-matching equation~(\ref{eq:deltak}) to have solutions, consistent with
energy conservation~(\ref{eq:energy}), for a pump centred at $800$~nm. We
solve~(\ref{eq:deltak}) for each process using the COMSOL effective indices
(figure~\ref{fig:refrac_index}) of the modes. The phase mismatch $\Delta k$ depends
on the pump frequency $f_p$ and on the offset $\delta$ of each photon from the pump,
energy conservation placing the two photons at $f_p \pm \delta$. Rather than solve
pair by pair, we plot every point where $\Delta k = 0$ while sweeping $f_p$ and
$\delta$, showing all phase-matched pairs at once as the curve in the
$(f_p, \delta)$ plane of figure~\ref{fig:phasematching_grid}. Reading a solution
then amounts to drawing a vertical line at the pump frequency
$f_p \approx 0.375\times10^{15}$~Hz ($\lambda_p = 800$~nm): each intersection is a
phase-matched pair, with $\delta$ read off the vertical axis,
$f_\mathrm{signal} = f_p + \delta$, $f_\mathrm{idler} = f_p - \delta$, and the
wavelengths following from $\lambda = c/f$. Only 15 processes admit such solutions,
those whose wavelengths appear in table~\ref{tab:combinations}.

Of these 15, combinations 23 and 49 admit several phase-matching solutions and are
discarded, since the spectral analysis reveals no signal peak beyond those of the
two detected pairs. As found in Section~\ref{sec:pumpdep}, the $648/1048$~nm pair is
generated by an orthogonal pump and the $660/1021$~nm pair by a co-polarised one,
and the pump polarisation of each process can likewise be read from its modes
(table~\ref{tab:combinations}) as co-polarised ($xx$ or $yy$) or orthogonal ($xy$);
matching each process to the detected pair of the same pump type attaches it to one
experimental pair. This attachment takes precedence over spectral proximity:
combination~11, for instance, predicts a pair at $647.6/1046.3$~nm, within $2$~nm of
the $648/1048$~nm pair, but its co-polarised pump excludes it from that pair, which
requires an orthogonal pump, and attaches it to the $660/1021$~nm pair. We then keep only the processes whose signal wavelength lies
within about $5\%$ of the detected pair of matching pump polarisation, leaving four
candidates, combinations 11, 21, 32 and 51, highlighted in
table~\ref{tab:combinations}. Their phase-matching curves are shown in
figure~\ref{fig:phasematching_grid}, where the intersection with the pump vertical
gives, for each, the signal and idler wavelengths of table~\ref{tab:combinations}.

These four processes are therefore the only candidates compatible with the
wavelengths of the two pairs generated by the microcoupler and characterised in
Section~\ref{sec:experiment}. Their predicted wavelengths nevertheless deviate from the
measured ones by up to $4.5\%$, the largest deviation being that of the idler of
combination~32 ($1095.2$~nm predicted for $1048$~nm measured). Such deviations reflect
the approximations of the model: the axis ratio is measured away from the waist and
assumed constant along the central region, the waist diameter is read from a side view
only, the guide is treated as longitudinally uniform, and the Kerr contribution to the
phase mismatch is neglected. The absolute predicted wavelengths therefore serve only to
pre-select candidates and are not used to discriminate between them. The
wavelength-deviation criterion alone does not single out the process behind each pair,
so we now bring in tomography and a closer study of the phase matching.

\subsection{Selection by polarisation and phase matching}
\label{sec:selection}

The pump-polarisation dependence links each candidate to a pair according to
whether its pump is co-polarised or orthogonal, while tomography adds an
independent constraint by comparing the signal polarisations with one another and
the idler polarisations with one another. The pump argument sorts the four
candidates into two groups: combinations 11 and 51, with a co-polarised ($xx$)
pump, go with the $660/1021$~nm pair, and combinations 21 and 32, with an
orthogonal ($xy$) pump, with the $648/1048$~nm pair, giving two candidates per
pair. It cannot go further, however: within each group both candidates share the
same pump polarisation, so the pump-dependence measurement returns the same
signature for the two and cannot separate them.

Tomography (Section~\ref{sec:tomography}) showed the two signal photons to have
similar output polarisations and the two idlers nearly orthogonal ones. A viable pair of
combinations must therefore draw one member from the co-polarised group and one
from the orthogonal group, and must also satisfy this constraint, with the signal
and idler polarisations, taken as the $x$ and $y$ of the simulated modes,
respectively similar and nearly orthogonal. Combination 11 (signal $x$, idler $x$),
for instance, cannot be paired with combination 21 (signal $y$, idler $x$), whose
signals would be orthogonal and contradict the polarisation analysis, but it can be
paired with combination 32 (signal $x$, idler $y$), consistent with the
measurements, so $(11,32)$ is a possible assignment. Applying the same reasoning to
the others leaves two assignments compatible with the relative polarisation
measurements, A~$(11,32)$ and B~$(51,21)$, whose properties are collected in
table~\ref{tab:candidates}. Absolute polarisation measurements would identify the
right one, since table~\ref{tab:candidates} shows that A predicts signal photons of
absolute polarisation $x$ whereas B predicts $y$, but this is beyond reach: the
fibres apply an unknown, wavelength-dependent polarisation transformation, so only
photons of similar wavelength, which we assume undergo nearly the same transformation
(Section~\ref{sec:tomography}), can be compared. As polarisation cannot decide between A and B, we turn to their phase
matching.

\begin{table}[htpb]
\centering
\caption{The two candidate assignments consistent with both polarisation
measurements. Each assignment attributes one combination to each detected pair:
the co-polarised pump to 660/1021~nm, the orthogonal pump to 648/1048~nm.
$\lambda_{\text{signal}}$ and $\lambda_{\text{idler}}$ are the phase-matched
wavelengths.}
\label{tab:candidates}
\vspace{6pt}
\begin{tabular}{l c r@{$\,+\,$}l c r@{$\,+\,$}l c r r}
\hline
\multicolumn{1}{c}{detected pair} & \multicolumn{1}{c}{comb.} & \multicolumn{2}{c}{pump} & & \multicolumn{2}{c}{signal $+$ idler} & \multicolumn{1}{c}{pol}
 & \multicolumn{1}{c}{$\lambda_{\text{signal}}$} & \multicolumn{1}{c}{$\lambda_{\text{idler}}$}\\
 & & \multicolumn{2}{c}{} & & \multicolumn{2}{c}{} & & \multicolumn{1}{c}{(nm)} & \multicolumn{1}{c}{(nm)}\\
\hline
\multicolumn{10}{l}{\emph{Assignment A}}\\
660/1021~nm & 11 & $\mathrm{LP}_{11}^{x\text{-}\mathrm{odd}}$ & $\mathrm{LP}_{11}^{x\text{-}\mathrm{odd}}$ & $\rightarrow$ & $\mathrm{LP}_{11}^{x\text{-}\mathrm{odd}}$ & $\mathrm{LP}_{11}^{x\text{-}\mathrm{odd}}$ & $xx\!\to\!x+x$ & 647.6 & 1046.3\\
648/1048~nm & 32 & $\mathrm{LP}_{11}^{y\text{-}\mathrm{odd}}$ & $\mathrm{LP}_{01}^{x}$ & $\rightarrow$ & $\mathrm{LP}_{01}^{x}$ & $\mathrm{LP}_{11}^{y\text{-}\mathrm{odd}}$ & $xy\!\to\!x+y$ & 630.2 & 1095.2\\[6pt]
\multicolumn{10}{l}{\emph{Assignment B}}\\
660/1021~nm & 51$^{*}$ & $\mathrm{LP}_{01}^{x}$ & $\mathrm{LP}_{11}^{x\text{-}\mathrm{odd}}$ & $\rightarrow$ & $\mathrm{LP}_{11}^{y\text{-}\mathrm{odd}}$ & $\mathrm{LP}_{01}^{y}$ & $xx\!\to\!y+y$ & 656.5 & 1023.8\\
648/1048~nm & 21$^{*}$ & $\mathrm{LP}_{11}^{y\text{-}\mathrm{odd}}$ & $\mathrm{LP}_{11}^{x\text{-}\mathrm{odd}}$ & $\rightarrow$ & $\mathrm{LP}_{11}^{y\text{-}\mathrm{odd}}$ & $\mathrm{LP}_{11}^{x\text{-}\mathrm{odd}}$ & $xy\!\to\!y+x$ & 658.0 & 1020.1\\
\hline
\multicolumn{10}{l}{\footnotesize $^{*}$daughters reordered from notebook order to place the signal ($\lambda_{\min}$) first.}
\end{tabular}
\end{table}

\begin{figure}[htbp]
\centering
\includegraphics[width=0.8\textwidth]{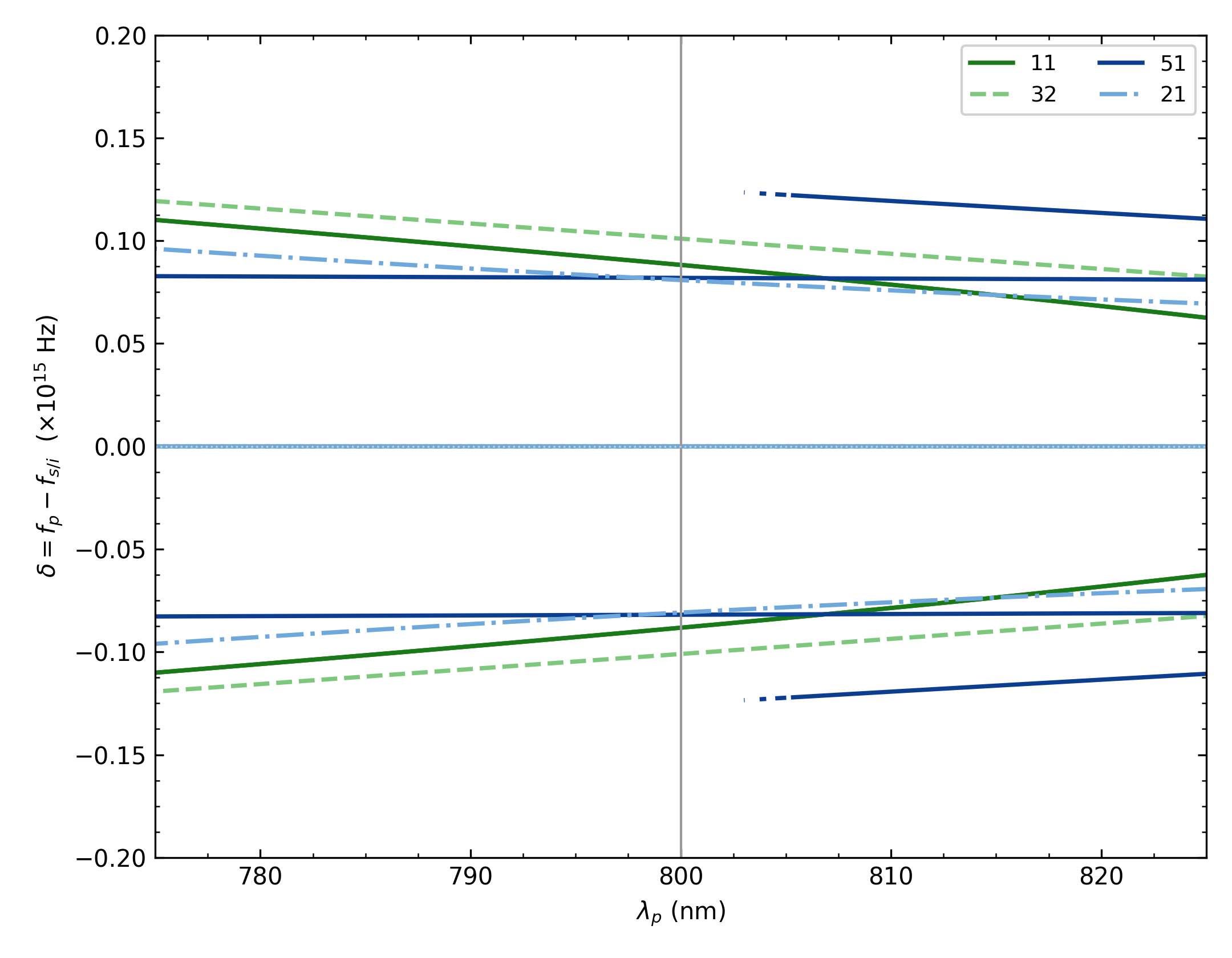}
\caption{Phase-matching branches of the four candidates near the experimental pump,
plotted as the photon--pump offset $\delta$ versus pump wavelength $\lambda_p$
(11 and 32 in green, 51 and 21 in blue). Around the pump, the two processes of
assignment B (51, 21) stay closer together than the two detected pairs, whereas
those of assignment A (11, 32) stay farther apart; at the experimental pump
(vertical line) the two B branches nearly coincide.}
\label{fig:phasematching_overlay}
\end{figure}

Taken process by process, the wavelengths would slightly favour assignment B, whose largest deviation from the measured pairs is $2.7\%$ against $4.5\%$ for A (table~\ref{tab:candidates}). The two assignments differ, however, in the separation between their two processes. The candidate table~\ref{tab:candidates}, which presents the candidate couples, shows that combinations 51 and 21 produce signals only 1.5~nm apart (656.5~nm and 658.0~nm), whereas the two detected pairs have signals more than 10~nm apart, so candidate B's two pairs would be hard to separate. The overlaid phase-matching curves (figure~\ref{fig:phasematching_overlay}) reinforce this: over a wavelength range centred on the pump, the gap between B's two processes stays below the gap between the observed pairs and drops to zero
within the pump bandwidth, whereas A's stays above. B's two processes would then
be too close to resolve, merging into a single pair, contradicting the observation
of two distinct pairs from different processes. This argument rests on the relative
position of two phase-matching curves rather than on absolute wavelengths, and thus
assumes that the uncertainties of the model, geometrical and nonlinear, shift the two
processes of B together. Discarding B in this way does not by itself establish A;
within the present geometrical model, the combined observations nevertheless favour
assignment A, the only one compatible with all the constraints.

Tomography provides an independent check of this conclusion. The hybrid mode
$\mathrm{LP}_{11}^{x\text{-}\mathrm{odd}}$ is not purely $x$-polarised: its $E_x$
component is about $2.5$ times its $E_y$ (Section~\ref{sec:geometry}), corresponding
to a linear polarisation at about $22^{\circ}$ from $x$. Assuming this ratio holds
across the mode profile and the polarisation is preserved through the transition
region, both assignments predict a trace distance of about $0.93$ between the idlers,
close to the measured $0.84\pm0.05$, but they differ for the signals. Assignment A,
whose signals lie in $\mathrm{LP}_{11}^{x\text{-}\mathrm{odd}}$ and
$\mathrm{LP}_{01}^{x}$, predicts about $0.37$, close to the measured $0.35\pm0.03$,
whereas assignment B, whose two signals share the $\mathrm{LP}_{11}^{y\text{-}\mathrm{odd}}$
mode, predicts $0$, so the measured value would have to come entirely from a
differential transformation in the output fibres. These predictions assume pure states;
the partial mixedness of the measured photons (purities between $0.83$ and $0.91$)
lowers the measured trace distances, consistent with the idler value of $0.84$ rather
than $0.93$. This comparison supports A without, on its own, excluding B.

%% file: section5_outlook.tex
\section{Outlook: polarisation entanglement}
\label{sec:outlook}

Since identifying the processes responsible for the experimentally observed photon pairs
revealed that two processes may emit photons nearly degenerate in wavelength under a single
pump wavelength, as is the case for $(51,21)$, it is natural to extend the theoretical study
of the device and ask what else it could produce, and in particular whether certain couples of
processes reach exact degeneracy, a necessary but not sufficient condition for generating
polarisation-entangled photon pairs.

Beyond the processes identified in Sec.~\ref{sec:modelling}, which generate photons close to
those observed experimentally, treating the pump wavelength as a free parameter rather than
fixing it at its experimental value allows a more systematic study of the couples of processes
that could occur simultaneously at a pump wavelength for which the signal and idler photons
are degenerate. Throughout this section the device is idealised, the six modes being assumed
perfectly resolved in polarisation; the actual limitation, which stems from the imperfectly
resolved hybrid modes of the present device (Sec.~\ref{sec:modelling}), is discussed at the end
of the section. In SFWM, two pump photons yield a signal--idler pair, and each process
corresponds to a choice of modes for these photons. Of all the possible processes, 51 are
allowed by the selection rules. Generating entangled photons requires the processes to combine
in pairs, so that with 51 allowed processes, 1275 combinations ($51\times50/2$) must be
examined. Retaining only those able to produce an entangled state reduces this number to 993,
which fall into three categories: 692 composite polarisation--spatial entangled combinations
(e.g. $(yo,yo)+(xe,xe) \rightarrow \lvert y,o;y,o\rangle+\lvert x,e;x,e\rangle$), 212 spatially
entangled (e.g. $(yo,yo)+(ye,ye) \rightarrow \lvert oo\rangle+\lvert ee\rangle$), and 89
polarisation entangled (e.g. $(xo,xo)+(yo,yo) \rightarrow \lvert xx\rangle+\lvert yy\rangle$).

We retain polarisation entanglement, as it is the only one directly measurable with standard
optical components, whereas spatial or hybrid entanglement would require a modal analysis. A
second selection is then applied to the 89 polarisation combinations. We discard those that
yield no phase-matching solution, those that are phase-matched but whose signal or idler
photons are not degenerate with their counterpart, and those whose phase matching produces
daughter photons close to the edges of the simulated wavelength range ($300$--$1200$~nm),
where the numerical accuracy is less reliable. The candidate studied below is chosen among
these 89 combinations and satisfies these three conditions. Related processes have been
studied in birefringent few-mode fibres~\cite{majchrowska2022,gawlik2025}; here the pairs are
produced in a fused microcoupler, for which generation and detection at the single-photon
level have already been established experimentally (Sec.~\ref{sec:experiment}).

\begin{figure}[htbp]
\centering
\includegraphics[width=\textwidth]{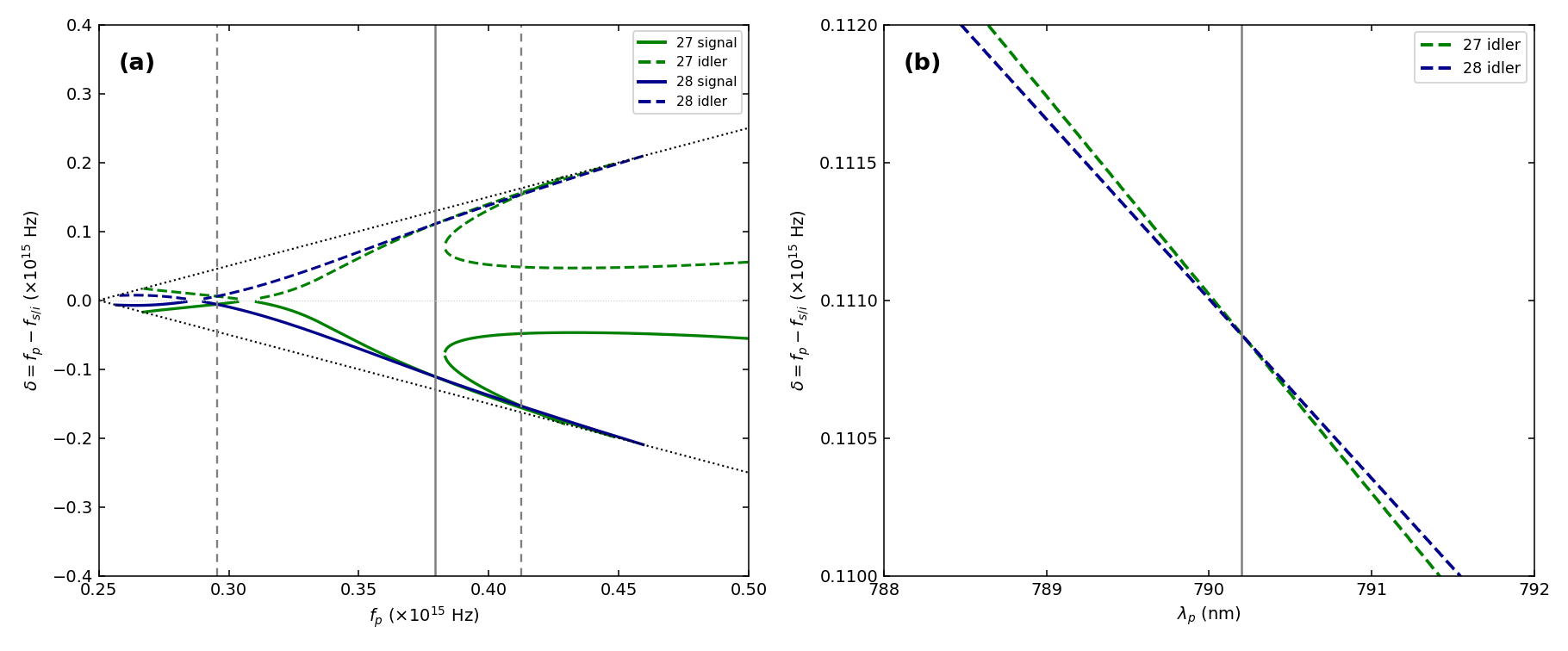}
\caption{Phase matching of the process pair $(27,28)$, plotted as the photon--pump
offset $\delta = f_p - f_{s/i}$ versus pump. (a)~Signal (solid) and idler (dashed)
branches of processes 27 (green) and 28 (blue) over the pump range
$f_p = 0.25$ to $0.5\times10^{15}$~Hz; the dotted diamond is the phase-matching
domain set by the $300$ to $1200$~nm simulation window. The solid vertical line
marks the polarisation-entangling crossing at $790.2$~nm, the two dashed vertical
lines the composite crossings at $726.9$ and $1015$~nm. (b)~Zoom on the $790.2$~nm
crossing, showing that the two idler branches coincide. Note: the plot shows only
the signal and idler wavelengths, not their spatial modes; across its turning point
process 27 exchanges the spatial modes of its signal and idler, whereas process 28
keeps them, which is what changes the entanglement type between the crossings.}
\label{fig:intrication_27_28}
\end{figure}

Among the couples producing polarisation entanglement, we retain one whose two processes share
exactly the same pump and keep the signal and the idler within the same LP$_{11}$ spatial
order, unlike other couples in which signal and idler belong to different orders. Its pump is
moreover accessible to a Ti:sapphire source. The selected couple is $(27,28)$, whose processes
are
$\lvert \mathrm{LP}_{11}^{x\text{-even}}\rangle + \lvert \mathrm{LP}_{11}^{x\text{-odd}}\rangle
\rightarrow
\lvert \mathrm{LP}_{11}^{x\text{-even}}\rangle + \lvert \mathrm{LP}_{11}^{x\text{-odd}}\rangle$
for 27 and
$\lvert \mathrm{LP}_{11}^{x\text{-even}}\rangle + \lvert \mathrm{LP}_{11}^{x\text{-odd}}\rangle
\rightarrow
\lvert \mathrm{LP}_{11}^{y\text{-even}}\rangle + \lvert \mathrm{LP}_{11}^{y\text{-odd}}\rangle$
for 28, the signal and idler photons lying in the same mode, LP$_{11}$-even and
LP$_{11}$-odd respectively. As the contour plots of figure~\ref{fig:intrication_27_28} show,
for a common pump at $790.2$~nm ($f_p = 0.3794\times10^{15}$~Hz) the signal photons are
produced at the same wavelength of $611.5$~nm
($|\delta| = |f_p - f_{s/i}| = 0.1109\times10^{15}$~Hz) and the idlers at $1116.5$~nm. At this
pump, the possible presence of a third phase-matched process, liable to generate spurious
pairs, would remain to be confirmed. The two-photon state $\lvert\psi\rangle$ of the signal
and idler photons can be written as
\begin{equation}
\lvert\psi\rangle \propto
a\,\lvert \mathrm{LP}_{11}^{x\text{-even}}\rangle_s \lvert \mathrm{LP}_{11}^{x\text{-odd}}\rangle_i
+ b\,e^{i\theta}\lvert \mathrm{LP}_{11}^{y\text{-even}}\rangle_s \lvert \mathrm{LP}_{11}^{y\text{-odd}}\rangle_i
\end{equation}
and, factorising the spatial part of the mode, one obtains
\begin{equation}
\lvert\psi\rangle \propto
\lvert \mathrm{even}\rangle_s \lvert \mathrm{odd}\rangle_i \otimes
\left( a\,\lvert x\rangle_s \lvert x\rangle_i + b\,e^{i\theta}\lvert y\rangle_s \lvert y\rangle_i \right),
\label{eq:etat_2_phton_xxyy}
\end{equation}
which shows that the spatial part is factorisable whereas the polarisation part is entangled.
Once the spectral and spatial degrees of freedom are traced out, the reduced polarisation
state reads
\begin{equation}
\rho_\mathrm{pol}\propto|a|^2\lvert xx\rangle\langle xx\rvert+|b|^2\lvert yy\rangle\langle yy\rvert
+\left(ab^{*}e^{-i\theta}\mu\,\lvert xx\rangle\langle yy\rvert+\mathrm{H.c.}\right),
\label{eq:rho_pol}
\end{equation}
where $\mu$ is the overlap between the spectral, temporal and spatial wave functions of the
two processes; a Bell state requires $|a|=|b|$, $|\mu|\approx1$ and a stable $\theta$. The two
amplitudes are not equal, however. Both processes share the same pump modes, so that the
excitation ratio of these modes affects $a$ and $b$ equally, and, for perfectly resolved
modes, the same spatial overlap integral, but process~27 involves $\chi_{XXXX}$ whereas
process~28 involves $\chi_{XXYY}$, which equals $\chi_{XXXX}/3$ for the electronic response
of silica~\cite{boyd}, so that $|b/a|=\Gamma_{28}/\Gamma_{27}\approx1/3$. For $\theta=0$ and
a real overlap $\mu$, and with normalised amplitudes, the fidelity to the Bell state
$\lvert \Phi^{+}\rangle$~\cite{nielsenchuang} is then $\tfrac{1}{2}+|ab|\,\mu\approx0.5+0.3\,\mu$
and the concurrence $2|ab||\mu|\approx0.6\,|\mu|$, that is, at most $0.80$ and $0.60$ even with
perfect indistinguishability: the state is entangled, but not maximally.

Figure~\ref{fig:intrication_27_28} further reveals two other possible crossings for the
process combination $(27,28)$: the first, at $\lambda_p = 726.9$~nm
($f_p = 0.412\times10^{15}$~Hz), produces signal photons at $529.7$~nm
($|\delta| = |f_p - f_{s/i}| = 0.1533\times10^{15}$~Hz) and an idler at $1157.8$~nm; the
second, at $\lambda_p = 1015$~nm, produces its daughter photons at $995.4$~nm
($|\delta| = |f_p - f_{s/i}| = 0.0060\times10^{15}$~Hz) and $1035.3$~nm. Note that this last
combination produces daughter photons close to the pump wavelength and would be difficult to
resolve. Both crossings produce the same type of two-photon state
\begin{equation}
\lvert\psi\rangle \propto
\lvert \mathrm{LP}_{11}^{x\text{-odd}}\rangle_s \lvert \mathrm{LP}_{11}^{x\text{-even}}\rangle_i
+ \lvert \mathrm{LP}_{11}^{y\text{-even}}\rangle_s \lvert \mathrm{LP}_{11}^{y\text{-odd}}\rangle_i
\end{equation}
which may be rewritten, using the same convention as
equation~\eqref{eq:etat_2_phton_xxyy}, as
\begin{equation}
\lvert\psi\rangle \propto
\lvert \mathrm{odd},x\rangle_s \lvert \mathrm{even},x\rangle_i
+ \lvert \mathrm{even},y\rangle_s \lvert \mathrm{odd},y\rangle_i .
\end{equation}
Here the signal photon has a different parity depending on the process, odd for 27 and even
for 28, so that parity is no longer common to the two terms and can no longer be factorised.
In contrast to the crossing at $790.2$~nm, which could be cast as the product of a
factorisable spatial state and an entangled polarisation state, the two-photon state at the
other two crossings takes the form of a single non-factorisable state, entangled in a
composite manner between spatial parity and polarisation, with the same amplitude imbalance
as at $790.2$~nm. Pumping the microcoupler so as to
address the same pair combination therefore yields different forms of entanglement depending
on the wavelength, in agreement with the tuning of entanglement by spectral overlap reported
in birefringent few-mode fibres~\cite{gawlik2025}.

To address the two processes generating the state of equation~\eqref{eq:etat_2_phton_xxyy},
the pump must simultaneously populate the LP$_{11}$ even and odd modes at the waist, both
polarised along $x$. An off-centre injection of the pump would allow these higher-order modes
to be excited in the central region~\cite{bilodeau1987}. The very structure of the
microcoupler, however, formed of two fused single-mode fibres, offers a natural way of
addressing these modes. Injecting the pump into a single arm excites several modes at the
waist, since the field of one core is not an eigenmode of the fused section and is therefore
projected onto several of them. Pumping both arms with a controlled relative phase, by
contrast, would make it possible to master this projection and select the desired mode or
superposition of modes. Such control of the LP$_{11}$ modes through the relative phase of the
two arms rests on mode-excitation and mode-shaping techniques already demonstrated in
fibre~\cite{ramachandrankristensen2013,bozinovic2013}.

At the crossing for a pump at $\lambda_p = 790.2$~nm, the spatial mode does not intrinsically
degrade the polarisation entanglement, since the signal and the idler are respectively even
and odd for both combinations, which allows the spatial part to be factorised, as
equation~\eqref{eq:etat_2_phton_xxyy} shows. Measuring the parity of a photon identifies only
whether it is the signal or the idler, not the process that generated it, so that the spatial
part carries no information distinguishing the two processes and does not degrade the
entanglement. The output fibres of the device, single-mode at the wavelengths of the daughter
photons (Sec.~\ref{sec:device}), project the LP$_{11}$ spatial structure onto their
fundamental mode; as this projection is a priori identical for both processes, it should
neither distinguish 27 from 28 nor degrade the polarisation entanglement, although this
remains to be quantified. The spectral signature, on the other hand, could reveal the process
that generated a photon pair, thereby degrading the entanglement. With a femtosecond pump,
processes 27 and 28 each generate photons within a band of finite
width~\cite{garaypalmett2007}; should these bands not overlap perfectly, or should Raman
scattering add uncorrelated photons in the same modes, the spectrum may betray the process and
degrade the entanglement. Beyond the amplitude ratio discussed above, the quality of the
entanglement is thus set by this spectral overlap, which enters $|\mu|$ in
equation~\eqref{eq:rho_pol},
and improving it requires tailoring the pump bandwidth, the dispersion, and the separation of
the bands from the Raman region~\cite{gawlik2025}. The
crossing therefore provides a candidate operating point for polarisation entanglement; a Bell
state would require balanced nonlinear amplitudes, a stable relative phase, and high spectral,
temporal and spatial indistinguishability.

This study thus suggests that entangled states of different types could be generated depending
on the operating point of the microcoupler. The processes involved, however, call upon hybrid
modes that are imperfectly resolved in polarisation on account of the moderate ellipticity of
the waist. This is a fabrication limit of the present device rather than a limit of principle:
a stronger ellipticity would resolve the six modes into well-defined
eigenstates~\cite{wang2005}, and the same combination $(27,28)$ would then generate a
polarisation-entangled state, although not a maximally entangled one owing to the imbalance
between its two nonlinear amplitudes. Future work could address the design of a new geometry, its
fabrication, its characterisation and the modal control of the pump, in order to demonstrate
the generation of a polarisation-entangled state.

%% file: section6_conclusion.tex
\section{Conclusion}
\label{sec:conclusion}

This work makes four contributions: it reports SFWM photon pairs in a multimode fused
microcoupler, introduces an identification method combining selection rules,
phase-matching calculations and two independent polarisation measurements, applies this
method to attribute the observed pairs to specific processes, and predicts theoretically
simultaneous processes able to generate entanglement. Within the present geometrical model,
we have identified the two spontaneous four-wave mixing processes, occurring in the waist of
the microcoupler, that give rise to the pairs observed experimentally at $648/1048$~nm and
$660/1021$~nm when the microcoupler is pumped at $800$~nm. Wavelength alone does not directly
identify the processes behind these pairs. The dependence of the coincidence rate between the
signal and the idler of a given pair on the pump polarisation, together with the tomography of
all the photons, which constitute two independent experimental signatures, do however reduce
the candidates to two couples without separating them, since the absolute polarisation of each
photon is not accessible. A closer analysis of the phase matching then favours
combination~11 for the $660/1021$~nm pair, whose pump photons are co-polarised ($xx$), and
combination~32 for the $648/1048$~nm pair, whose pump photons are orthogonal ($xy$), which
requires a pump polarised at $45^{\circ}$. In this assignment, the two signal photons are both
$x$-polarised at the waist whereas the idler photons are orthogonal ($x$ and $y$), consistent
with the similar and nearly orthogonal output polarisations measured by tomography. The first
process~(11) is purely intramodal, all the modes involved lying in the hybrid
$\mathrm{LP}_{11}^{x\text{-}\mathrm{odd}}$ mode, whereas the second is intermodal since it
couples a fundamental $\mathrm{LP}_{01}$ mode to an $\mathrm{LP}_{11}$ mode.

The geometry on which the numerical simulations are based is derived from SEM images of finite
precision, and the predicted wavelengths deviate from the measured ones by up to $4.5\%$. The
polarisation measurements, which carry most of the identification, do not depend on this
geometry, although the tomographic comparison assumes that photons of similar wavelength
undergo nearly the same transformation in the output fibres. The phase matching that discards
the last candidate couple does rely on simulated modes, but it enters only through a ratio:
this couple predicts two pairs too close in wavelength to be distinguished, whereas the
experiment resolves two. Establishing that model errors, geometrical or nonlinear, cannot
bridge this gap would require a systematic sensitivity analysis of the model; within the
present model, the retained assignment is the only one consistent with all the constraints.
In our multimode device, the identification rests on the crossing of several independent
experimental signatures, spectral and polarimetric, supported by a numerical simulation. None
suffices on its own. The spectrum identifies several candidates within a few percent of the
observed wavelengths. The pump-polarisation dependence, together with the spectral analysis,
attaches a candidate theoretical process to one of the experimentally observed pairs.
Tomography, which constrains the relative polarisations of the daughter photons, leaves only
two couples of processes that a closer analysis of the phase matching separates.

With this identification complete, a natural extension of the study is to predict
theoretically that a couple of processes sharing a pump in the same spatial mode, for instance
$(27,28)$, could emit an entangled state whose type of entanglement, polarisation or composite,
depends on the common pump wavelength. This identifies a candidate operating point for
generating entanglement with such a device. Its realisation nevertheless remains limited by
the moderate ellipticity of the waist, which leaves two of the four $\mathrm{LP}_{11}$ modes
only partially resolved. This is a fabrication limit rather than one of principle, since a
stronger ellipticity of the core would resolve the issue. A second limit is tied to the
spectral overlap of the two processes, which the use of a femtosecond pump would require
controlling, and a third to the imbalance between their nonlinear amplitudes, which leaves the
state only partially entangled.

As a complete characterisation of a complex device, combining experiment and simulation, our
work shows how the full potential of such a device could be reached. A new, more strongly
elliptical geometry would remain to be designed and fabricated, then characterised following
the method established in this study. The control of the spatial mode of the pump would remain
to be demonstrated. By pumping both arms of such a device and controlling the relative phase
between them, it becomes conceivable to select a specific spatial mode of the pump and, in
principle, to eliminate the multimode contamination.

%% file: references.bib
@article{baker2011,
  author  = {Baker, C. and Rochette, M.},
  title    = {A generalized heat-brush approach for precise control of the waist profile in fiber tapers},
  journal  = {Optical Materials Express},
  volume   = {1},
  number   = {6},
  pages    = {1065--1076},
  year     = {2011},
  doi      = {10.1364/OME.1.001065}
}

@article{birks1992,
  author  = {Birks, T. A. and Li, Y. W.},
  title    = {The shape of fiber tapers},
  journal  = {Journal of Lightwave Technology},
  volume   = {10},
  number   = {4},
  pages    = {432--438},
  year     = {1992},
  doi      = {10.1109/50.134196}
}

@article{bilodeau1987,
  author  = {Bilodeau, F. and Hill, K. O. and Johnson, D. C. and Faucher, S.},
  title    = {Compact, low-loss, fused biconical taper couplers: overcoupled operation and antisymmetric supermode cutoff},
  journal  = {Optics Letters},
  volume   = {12},
  number   = {8},
  pages    = {634--636},
  year     = {1987},
  doi      = {10.1364/OL.12.000634}
}

@article{wang2005,
  author  = {Wang, Z. and Ju, J. and Jin, W.},
  title    = {Properties of elliptical-core two-mode fiber},
  journal  = {Optics Express},
  volume   = {13},
  number   = {11},
  pages    = {4350--4357},
  year     = {2005},
  doi      = {10.1364/OPEX.13.004350}
}

@book{keiser,
  author    = {Keiser, G.},
  title     = {Optical Fiber Communications},
  edition   = {4th},
  publisher = {McGraw-Hill},
  year      = {2011},
  note      = {Sec.~2.4}
}

@book{agrawal,
  author    = {Agrawal, G. P.},
  title     = {Nonlinear Fiber Optics},
  edition   = {6th},
  publisher = {Academic Press},
  year      = {2019},
  isbn      = {978-0-12-817042-7}
}

@book{loudon,
  author    = {Loudon, R.},
  title     = {The Quantum Theory of Light},
  edition   = {3rd},
  publisher = {Oxford University Press},
  year      = {2000},
  isbn      = {978-0-19-850177-0}
}

@book{nielsenchuang,
  author    = {Nielsen, M. A. and Chuang, I. L.},
  title     = {Quantum Computation and Quantum Information},
  edition   = {10th Anniversary},
  publisher = {Cambridge University Press},
  year      = {2010},
  note      = {Sec.~9.2 (trace distance)}
}

@article{gloge1971,
  author    = {Gloge, D.},
  title     = {Weakly Guiding Fibers},
  journal   = {Applied Optics},
  volume    = {10},
  number    = {10},
  pages     = {2252--2258},
  year      = {1971}
}

@article{kogelnikwinzer2012,
  author    = {Kogelnik, H. and Winzer, P. J.},
  title     = {Modal Birefringence in Weakly Guiding Fibers},
  journal   = {Journal of Lightwave Technology},
  volume    = {30},
  number    = {14},
  pages     = {2240--2245},
  year      = {2012},
  doi       = {10.1109/JLT.2012.2193872}
}

@book{boyd,
  author    = {Boyd, R. W.},
  title     = {Nonlinear Optics},
  edition   = {3rd},
  publisher = {Academic Press},
  year      = {2008},
  isbn      = {978-0-12-369470-6}
}

@article{garaypalmett2016,
  author = {Garay-Palmett, K. and Cruz-Delgado, D. and Dominguez-Serna, F. and Ortiz-Ricardo, E. and Monroy-Ruz, J. and Cruz-Ramirez, H. and Ramirez-Alarcon, R. and U'Ren, A. B.},
  title   = {Photon pair generation by intermodal spontaneous four-wave mixing in birefringent, weakly guiding optical fibers},
  journal = {Physical Review A},
  volume  = {93},
  number  = {3},
  pages   = {033810},
  year    = {2016},
  doi     = {10.1103/PhysRevA.93.033810},
}

@article{rottwitt2018,
  author  = {Rottwitt, K. and Koefoed, J. G. and Christensen, E. N.},
  title   = {Photon-pair sources based on intermodal four-wave mixing in few-mode fibers},
  journal = {Fibers},
  volume  = {6},
  number  = {2},
  pages   = {32},
  year    = {2018},
  doi     = {10.3390/fib6020032}
}

@article{bozinovic2013,
  author  = {Bozinovic, N. and Yue, Y. and Ren, Y. and Tur, M. and Kristensen, P. and Huang, H. and Willner, A. E. and Ramachandran, S.},
  title   = {Terabit-scale orbital angular momentum mode division multiplexing in fibers},
  journal = {Science},
  volume  = {340},
  number  = {6140},
  pages   = {1545--1548},
  year    = {2013},
  doi     = {10.1126/science.1237861}
}

@article{ramachandrankristensen2013,
  author  = {Ramachandran, S. and Kristensen, P.},
  title   = {Optical vortices in fiber},
  journal = {Nanophotonics},
  volume  = {2},
  number  = {5-6},
  pages   = {455--474},
  year    = {2013},
  doi     = {10.1515/nanoph-2013-0047}
}

@inproceedings{shahar2023,
  author    = {Shahar, D. I. and Liu, X. and Kim, D. B. and Lorenz, V. O. and Ramachandran, S.},
  title     = {Photon pair generation in {OAM} modes at 780 and 1550 nm via spontaneous intermodal four-wave mixing},
  booktitle = {CLEO: Fundamental Science},
  pages     = {FF1L.5},
  year      = {2023},
  organization = {Optica Publishing Group}
}

@article{james2001,
  author  = {James, D. F. V. and Kwiat, P. G. and Munro, W. J. and White, A. G.},
  title   = {Measurement of qubits},
  journal = {Physical Review A},
  volume  = {64},
  number  = {5},
  pages   = {052312},
  year    = {2001},
  doi     = {10.1103/PhysRevA.64.052312}
}

@article{tapster1998,
  author  = {Tapster, P. R. and Rarity, J. G.},
  title   = {Photon statistics of pulsed parametric light},
  journal = {Journal of Modern Optics},
  volume  = {45},
  number  = {3},
  pages   = {595--604},
  year    = {1998},
  doi     = {10.1080/09500349808231917}
}

@book{blackgagnon,
  author    = {Black, R. J. and Gagnon, L.},
  title     = {Optical Waveguide Modes: Polarization, Coupling and Symmetry},
  publisher = {McGraw-Hill},
  year      = {2010},
  isbn      = {978-0-07-162296-7}
}

@article{majchrowska2022,
  author  = {Majchrowska, S. and \.{Z}o{\l}nacz, K. and Urba\'{n}czyk, W. and Tarnowski, K.},
  title   = {Multiple intermodal-vectorial four-wave mixing bands generated by selective excitation of orthogonally polarized LP01 and LP11 modes in a birefringent fiber},
  journal = {Optics Letters},
  volume  = {47},
  number  = {10},
  pages   = {2522--2525},
  year    = {2022},
  doi     = {10.1364/OL.456521}
}

@article{gawlik2025,
  author  = {Gawlik, A. and Berna\'{s}, M. and \.{Z}o{\l}nacz, K. and Tarnowski, K.},
  title   = {Towards generation of hybrid entangled photon pairs by spectrally overlapping intermodal-vectorial four-wave mixing bands in optical fibers},
  journal = {Scientific Reports},
  volume  = {15},
  pages   = {45490},
  year    = {2025},
  doi     = {10.1038/s41598-025-29082-3}
}

@article{garaypalmett2007,
  author  = {Garay-Palmett, K. and McGuinness, H. J. and Cohen, O. and Lundeen, J. S. and Rangel-Rojo, R. and U'Ren, A. B. and Raymer, M. G. and McKinstrie, C. J. and Radic, S. and Walmsley, I. A.},
  title   = {Photon pair-state preparation with tailored spectral properties by spontaneous four-wave mixing in photonic-crystal fiber},
  journal = {Optics Express},
  volume  = {15},
  number  = {22},
  pages   = {14870--14886},
  year    = {2007},
  doi     = {10.1364/OE.15.014870}
}

@article{Bennink2010,
  author  = {Bennink, Ryan S.},
  title   = {Optimal collinear {G}aussian beams for spontaneous parametric down-conversion},
  journal = {Physical Review A},
  volume  = {81}, number = {5}, pages = {053805}, year = {2010},
  doi     = {10.1103/PhysRevA.81.053805}
}

@article{Guerreiro2013,
  author  = {Guerreiro, T. and Martin, A. and Sanguinetti, B. and Bruno, N. and Zbinden, H. and Thew, R. T.},
  title   = {High efficiency coupling of photon pairs in practice},
  journal = {Optics Express},
  volume  = {21}, number = {23}, pages = {27641--27651}, year = {2013},
  doi     = {10.1364/OE.21.027641}
}

@article{Anwar2021,
  author  = {Anwar, Ali and Perumangatt, Chithrabhanu and Steinlechner, Fabian and Jennewein, Thomas and Ling, Alexander},
  title   = {Entangled photon-pair sources based on three-wave mixing in bulk crystals},
  journal = {Review of Scientific Instruments},
  volume  = {92}, number = {4}, pages = {041101}, year = {2021},
  doi     = {10.1063/5.0023103}
}

@article{Martin2010,
  author  = {Martin, A. and Issautier, A. and Herrmann, H. and Sohler, W. and Ostrowsky, D. B. and Alibart, O. and Tanzilli, S.},
  title   = {A polarization entangled photon-pair source based on a type-II {PPLN} waveguide emitting at a telecom wavelength},
  journal = {New Journal of Physics},
  volume  = {12}, number = {10}, pages = {103005}, year = {2010},
  doi     = {10.1088/1367-2630/12/10/103005}
}

@article{GisinThew2007,
  author  = {Gisin, Nicolas and Thew, Rob},
  title   = {Quantum communication},
  journal = {Nature Photonics},
  volume  = {1}, number = {3}, pages = {165--171}, year = {2007},
  doi     = {10.1038/nphoton.2007.22}
}

@article{Pirandola2020,
  author  = {Pirandola, S. and Andersen, U. L. and Banchi, L. and Berta, M. and Bunandar, D. and Colbeck, R. and Englund, D. and Gehring, T. and Lupo, C. and Ottaviani, C. and Pereira, J. L. and Razavi, M. and Shaari, J. S. and Tomamichel, M. and Usenko, V. C. and Vallone, G. and Villoresi, P. and Wallden, P.},
  title   = {Advances in quantum cryptography},
  journal = {Advances in Optics and Photonics},
  volume  = {12}, number = {4}, pages = {1012--1236}, year = {2020},
  doi     = {10.1364/AOP.361502}
}

@article{OBrien2009,
  author  = {O'Brien, Jeremy L. and Furusawa, Akira and Vu\v{c}kovi\'{c}, Jelena},
  title   = {Photonic quantum technologies},
  journal = {Nature Photonics},
  volume  = {3}, number = {12}, pages = {687--695}, year = {2009},
  doi     = {10.1038/nphoton.2009.229}
}

@article{Flamini2019,
  author  = {Flamini, Fulvio and Spagnolo, Nicol\`{o} and Sciarrino, Fabio},
  title   = {Photonic quantum information processing: a review},
  journal = {Reports on Progress in Physics},
  volume  = {82}, number = {1}, pages = {016001}, year = {2019},
  doi     = {10.1088/1361-6633/aad5b2}
}

@article{Horodecki2009,
  author  = {Horodecki, Ryszard and Horodecki, Pawe\l{} and Horodecki, Micha\l{} and Horodecki, Karol},
  title   = {Quantum entanglement},
  journal = {Reviews of Modern Physics},
  volume  = {81}, number = {2}, pages = {865--942}, year = {2009},
  doi     = {10.1103/RevModPhys.81.865}
}

@article{GarayPalmett2023,
  author  = {Garay-Palmett, K. and Kim, D. B. and Zhang, Y. and Dom\'{i}nguez-Serna, F. A. and Lorenz, V. O. and U'Ren, A. B.},
  title   = {Fiber-based photon-pair generation: tutorial},
  journal = {Journal of the Optical Society of America B},
  volume  = {40}, number = {3}, pages = {469--490}, year = {2023},
  doi     = {10.1364/JOSAB.478008}
}

@article{Afsharnia2024,
  author  = {Afsharnia, Mina and Junaid, Saher and Saravi, Sina and Chemnitz, Mario and Wondraczek, Katrin and Pertsch, Thomas and Schmidt, Markus A. and Setzpfandt, Frank},
  title   = {Generation of infrared photon pairs by spontaneous four-wave mixing in a {CS$_2$}-filled microstructured optical fiber},
  journal = {Scientific Reports},
  volume  = {14}, pages = {977}, year = {2024},
  doi     = {10.1038/s41598-024-51482-0}
}

@article{CruzDelgado2016,
  author  = {Cruz-Delgado, D. and Ramirez-Alarcon, R. and Ortiz-Ricardo, E. and Monroy-Ruz, J. and Dominguez-Serna, F. and Cruz-Ramirez, H. and Garay-Palmett, K. and U'Ren, A. B.},
  title   = {Fiber-based photon-pair source capable of hybrid entanglement in frequency and transverse mode, controllably scalable to higher dimensions},
  journal = {Scientific Reports},
  volume  = {6}, pages = {27377}, year = {2016},
  doi     = {10.1038/srep27377}
}

@article{Montazeri2024,
  author  = {Montazeri, Seyedehnajmeh and Zobair, Md. Abu and Esmaeelpour, Mina},
  title   = {Investigation of intermodal four-wave mixing for continuous-wave photon-pair generation},
  journal = {IEEE Photonics Technology Letters},
  volume  = {36}, number = {9}, pages = {605--608}, year = {2024},
  doi     = {10.1109/LPT.2024.3382688}
}

@article{Shukhin2020,
  author  = {Shukhin, A. A. and Keloth, J. and Hakuta, K. and Kalachev, A. A.},
  title   = {Heralded single photon and correlated photon pair generation via spontaneous four-wave mixing in tapered optical fibers},
  journal = {Physical Review A},
  volume  = {101}, number = {5}, pages = {053822}, year = {2020},
  doi     = {10.1103/PhysRevA.101.053822}
}

@mastersthesis{Cheng2017,
  author  = {Cheng, Xinru},
  title   = {Generation of Photon Pairs in Fiber Microcouplers},
  school  = {University of Ottawa},
  address = {Ottawa, Canada},
  year    = {2017},
  doi     = {10.20381/ruor-604}
}
